\documentclass[
reprint,
superscriptaddress,
amsmath,amssymb,
aps,
pra]{revtex4-1}
\usepackage{amsmath,amssymb}
\usepackage{graphicx}
\usepackage{dcolumn}
\usepackage{bm}
\usepackage{physics}
\usepackage{braket}
\usepackage{amssymb}
\usepackage{placeins}
\usepackage{array}
\usepackage{lipsum}
\usepackage{color,soul}
\usepackage{upgreek}
\RequirePackage[colorlinks=true,linkcolor=blue,urlcolor=blue,hyperfootnotes=false,citecolor=black]{hyperref}

\begin{document}
\title{Observation and stabilization of photonic Fock states\\ in a hot radio-frequency resonator}

\author{Mario F. Gely}
\affiliation{%
Kavli Institute of NanoScience, Delft University of Technology,\\
PO Box 5046, 2600 GA, Delft, The Netherlands.
}

\author{Marios Kounalakis}
\affiliation{%
Kavli Institute of NanoScience, Delft University of Technology,\\
PO Box 5046, 2600 GA, Delft, The Netherlands.
}

\author{Christian Dickel}
\affiliation{%
Kavli Institute of NanoScience, Delft University of Technology,\\
PO Box 5046, 2600 GA, Delft, The Netherlands.
}

\author{Jacob Dalle}
\affiliation{%
Kavli Institute of NanoScience, Delft University of Technology,\\
PO Box 5046, 2600 GA, Delft, The Netherlands.
}

\author{R\'{e}my Vatr\'{e}}
\affiliation{%
Kavli Institute of NanoScience, Delft University of Technology,\\
PO Box 5046, 2600 GA, Delft, The Netherlands.
}

\author{Brian Baker}
\affiliation{%
Department of Physics and Astronomy, Northwestern University,\\
Evanston, Illinois 60208, USA.
}

\author{Mark D. Jenkins}
\affiliation{%
Kavli Institute of NanoScience, Delft University of Technology,\\
PO Box 5046, 2600 GA, Delft, The Netherlands.
}

\author{Gary A. Steele}
\affiliation{%
Kavli Institute of NanoScience, Delft University of Technology,\\
PO Box 5046, 2600 GA, Delft, The Netherlands.
}

\begin{abstract}
Detecting weak radio-frequency electromagnetic fields plays a crucial role in a wide range of fields, from radio astronomy to nuclear magnetic resonance imaging.
In quantum optics, the ultimate limit of a weak field is a single photon.
Detecting and manipulating single photons at megahertz frequencies presents a challenge as, even at cryogenic temperatures, thermal fluctuations are appreciable.
Using a gigahertz superconducting qubit, we observe the quantization of a megahertz radio-frequency resonator, cool it to the ground-state and stabilize Fock states.
Releasing the resonator from our control, we observe its re-thermalization with nanosecond resolution.
Extending circuit quantum electrodynamics to the megahertz regime, we enable the exploration of thermodynamics at the quantum scale and allow interfacing quantum circuits with megahertz systems such as spin systems or macroscopic mechanical oscillators.
\end{abstract}

\maketitle 

\let\oldaddcontentsline\addcontentsline
\renewcommand{\addcontentsline}[3]{}

\begin{figure*}[ht!]
\makebox[\textwidth][c]{\includegraphics[]{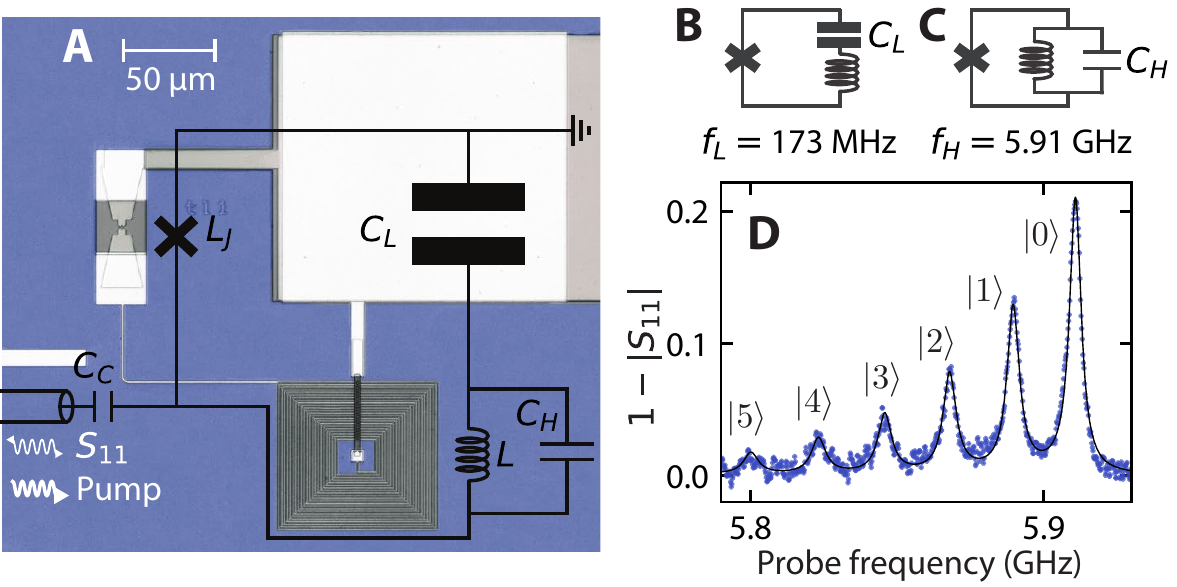}}
\caption[]{
\textbf{Cross-Kerr coupling between a transmon qubit and radio-frequency resonator. A: }
False-colored optical micrograph of the device overlaid with the equivalent lumped element circuit. 
\textbf{B,C: }
Effective circuit at low and high frequencies.
At low (high) frequencies, the femtofarad (picofarad) capacitances of the circuit are equivalent to open (short) circuits, and the device is equivalent to a series (parallel) JJ-inductor-capacitor combination.
The circuit has thus two modes, a $173$~MHz resonator and a $5.9$~GHz qubit.
\textbf{D: }
Microwave response $|S_{11}|$.
Through cross-Kerr coupling, quantum fluctuations of a photon number state $|n=0,1,..\rangle$ in the resonator shift the qubit transition frequency.
Peak heights are proportional to the occupation of state $|n\rangle$, and we extract a thermal occupation $n_\text{th} = 1.6$ in the resonator corresponding to a temperature of 17~mK. 
}
\end{figure*}

\begin{figure*}[ht!]
\makebox[\textwidth][c]{\includegraphics[]{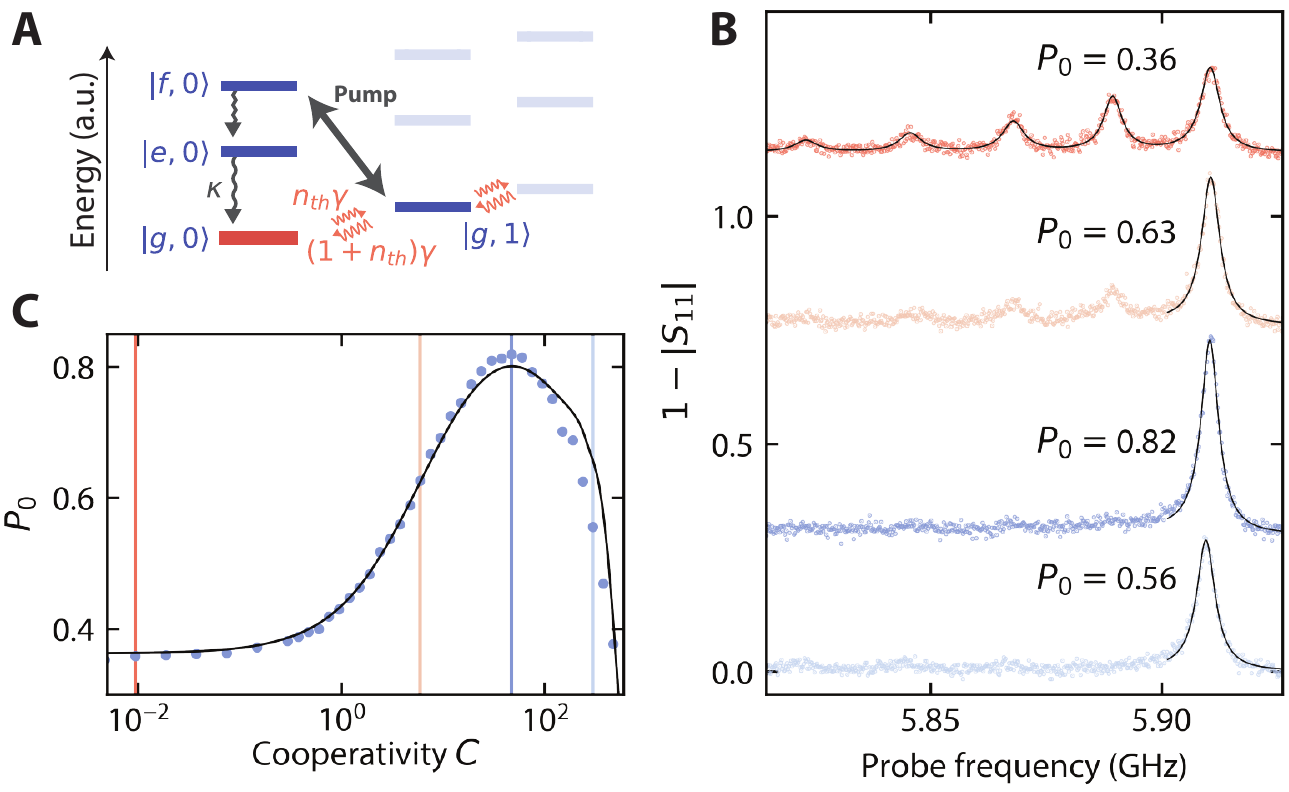}}
\caption[]{
\textbf{Ground-state cooling of the radio-frequency resonator. A: }
Energy ladder of the coupled transmon qubit and resonator.
Meandering arrows indicate relaxation and thermal processes. 
The resonator is cooled by driving a transition (black arrow) that transfers excitations from the resonator to the qubit, where they are quickly dissipated.
\textbf{B: }
Photon-number spectroscopy of the resonator for different cooperativities $C$ (proportional to cooling-pump power).
$C = 0.01,\ 6,\ 47,\ 300$ from top to bottom.
Ground-state occupations $P_0$ are extracted from Lorentzian fits (black curves). 
\textbf{C: }
%
%
Vertical lines indicate the the datasets of panel B.
A simulation (curve) predicts the measured (dots) high-$C$ decrease of $P_0$ through the off-resonant driving of other sideband transitions.
}
\end{figure*}

\begin{figure*}[ht!]
\makebox[\textwidth][c]{\includegraphics[]{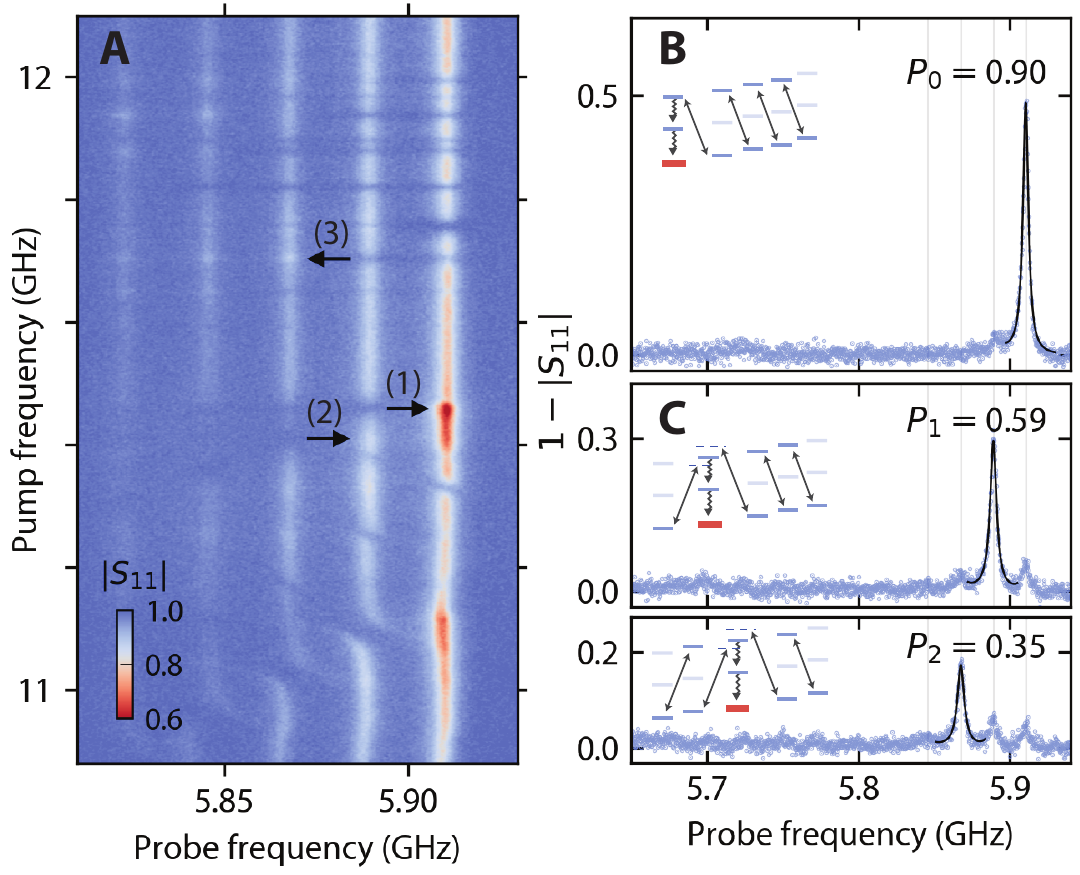}}
\caption[]{
\textbf{Enhanced cooling and Fock-state stabilization using multiple tones. }
\textbf{A:} $|S_{11}|$ as a function of pump and probe frequency.
Vertical lines correspond to the photon-number splitted qubit frequencies. 
Horizontal and diagonal features appear at pump frequencies enabling the transfer of population between Fock states of the resonator. 
Arrows indicate three example transitions:
\textbf{(1)} the cooling transition of Fig.~2,
\textbf{(2)} the transition $|g,2\rangle\leftrightarrow|f,1\rangle$ transferring $|2\rangle$ to $|1\rangle$, and
\textbf{(3)} the transition $|g,1\rangle\leftrightarrow|f,2\rangle$ which raises $|1\rangle$ to $|2\rangle$.
\textbf{B: } 
By simultaneously driving four cooling transitions ($|g,n+1\rangle\leftrightarrow|f,n\rangle$), cooling is enhanced to $P_0 = 0.9$. 
\textbf{C: } 
Using these transitions in conjunction with raising transitions $|g,n\rangle\leftrightarrow|f,n+1\rangle$, we stabilize Fock states 1 and 2 with $59\%$ and $35\%$ fidelity.
We fit a sum of complex Lorentzians to the spectrum, showing only the relevant Lorentzian (black curve) whose amplitude provides $P_n$.
Off-resonant driving results in population transfer to higher energy states visible as features in the lower frequencies of the spectrum.
}
\end{figure*}

\begin{figure}[ht!]
\makebox[0.5\textwidth][c]{\includegraphics[]{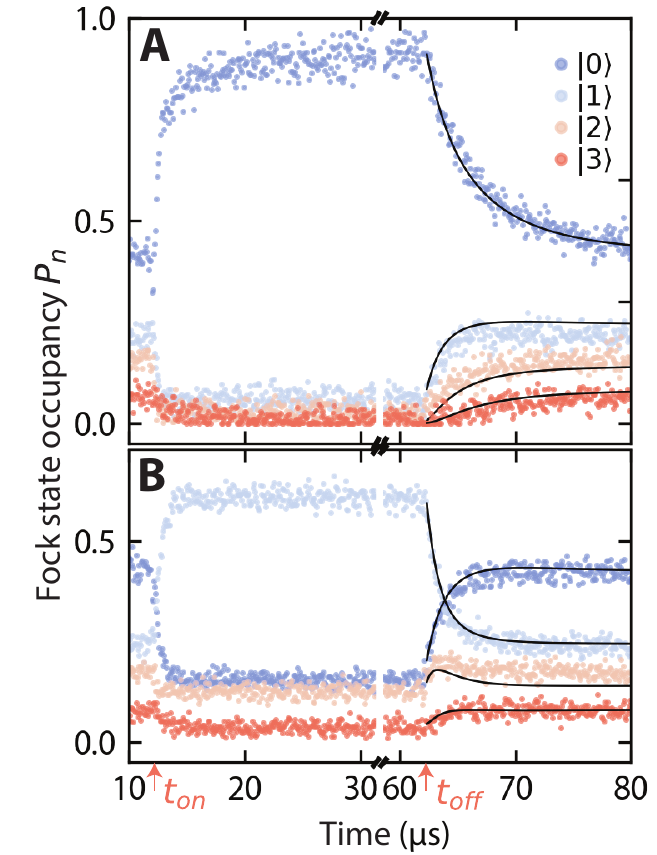}}
\caption[]{\textbf{Fock-state-resolved thermalization-dynamics of the resonator.}
At $t_\text{on}$, pumps are turned on and the resonator evolves into the ground-state (\textbf{A}) or a single photon state (\textbf{B}). 
At $t_\text{off}$, control is released and we observe photon-number resolved thermalization of the resonator. 
The extracted Fock-state occupation (dots) is fitted to Eq.~\ref{eq:main_rate_equation} (black curve). 
}
\end{figure}

Detecting and manipulating single photons becomes more difficult at lower frequencies because of thermal fluctuations.
A hot environment randomly creates and annihilates photons causing decoherence in addition to creating statistical mixtures of states from which quantum state preparation is challenging.
This can be mitigated by using a colder system as a heat sink, to extract the entropy created by the environment.
Such a scheme, known as reservoir engineering, was first developed in trapped ions~\cite{Diedrich1989}, where hot degrees of freedom are cooled via the atomic transitions of ions.

Using superconducting electronics, circuit quantum electrodynamics (cQED) has made extensive use of reservoir engineering to cool, but also manipulate electromagnetic fields at the quantum level.
With the prospect of building a quantum computer, or to demonstrate fundamental phenomena, experiments have shown the cooling or reset of qubits to their ground-state~\cite{Valenzuela2006,Geerlings2013,Magnard2018}, also in the megahertz regime~\cite{Vool2018}, quantum state stabilization~\cite{Murch2012,Shankar2013,Leghtas2015}, and quantum error correction~\cite{Ofek2016}.
Using superconducting circuits, reservoir engineering is commonplace in electromechanical systems~\cite{Teufel2011,Eichler2018}, but with weak non-linearity, such schemes have only limited quantum control~\cite{ockeloen2018stabilized,Viennot2018} compared to typical cQED systems.
Despite the many applications of quantum state engineering in cQED, obtaining control over the quantum state of a hot resonator, where the environment temperature is a dominant energy scale, remains a largely unexplored and challenging task.

We directly observe the quantization of radio-frequency electromagnetic fields in a thermally-excited megahertz photonic resonator, and manipulate its quantum state using reservoir engineering.
Specifically, we cool the 173~MHz resonator to $90\%$ ground-state occupation, and stabilize one- and two-photon Fock states. 
Releasing the resonator from our control, we observe its re-thermalization with photon-number resolution.
We use the paradigm of cQED, where a resonator can be read out and controlled by dispersively coupling it to a superconducting qubit.
Achieving this with a GHz qubit and MHz photons is challenging, since in a conventional cQED architecture the coupling would be far too weak~\cite{gely2017nature}.
To overcome this, we present a circuit enabling a very strong interaction, resulting in a  cross-Kerr coupling larger than the qubit and resonator dissipation rates, despite an order of magnitude difference in their resonance frequencies.
The circuit (Fig.~1A) comprises of a Josephson junction ($L_\text{J}=41$ nH) connected in series to a capacitor ($C_\text{L}=11$~pF) and a spiral inductor ($L=28$ nH).
At low frequencies, the parasitic capacitance of the spiral inductor is negligible, and the equivalent circuit (Fig.~1B) has a first transition frequency $\omega_\text{L} = 2\pi\times173\ \text{MHz}$.
At gigahertz frequencies, $C_L$ behaves as a short, and the capacitance of the spiral inductor $C_\text{H}=40$~fF becomes relevant instead.
The resulting parallel connection of $L_\text{J}$, $L$ and $C_\text{H}$ (Fig.~1C) has a first transition frequency $\omega_\text{H}=2\pi\times5.91\ \text{GHz}$.
The two modes share the Josephson junction.
The junction has an inductance that varies with the current fluctuations traversing it, and consequently the resonance frequency of the high-frequency (HF) mode shifts as a function of the number of excitations in the low-frequency (LF) mode and vice versa.
This cross-Kerr interaction is quantified by the shift per photon $\chi = 2\sqrt{A_\text{H}A_\text{L}}$, where the anharmonicity of the LF and HF mode $A_\text{L}= h\times495\ \text{kHz}$ and $A_\text{H}=h\times192\ \text{MHz}$ are given by~\cite{SI}
\begin{equation}
  A_\text{L} = -\frac{e^2}{2C_\text{L}}\left(\frac{L_\text{J}}{L+L_\text{J}}\right)^3,\ A_\text{H} = -\frac{e^2}{2C_\text{H}}\left(\frac{L}{L+L_\text{J}}\right).
  \label{eq:anharmonicities}
\end{equation}

The system is described by the Hamiltonian~\cite{SI}
\begin{equation}
\begin{split}
    \hat{H} &= \hbar\omega_\text{H}\hat{a}^\dagger\hat{a}+ \hbar\omega_\text{L}\hat{b}^\dagger\hat{b}\\
    &  -\frac{A_\text{H}}{2}\hat{a}^\dagger\hat{a}^\dagger\hat{a}\hat{a}-\frac{A_\text{L}}{2}\hat{b}^\dagger\hat{b}^\dagger\hat{b}\hat{b}\\
    &-\chi\hat{a}^\dagger\hat{a}\hat{b}^\dagger\hat{b}\ ,\\
    \label{eq:main_hamiltonian}
\end{split}
\end{equation}
where $\hat a$ ($\hat b$) is the annihilation operator for photons in the HF (LF) mode.
The second line describes the anharmonicity or Kerr non-linearity of each mode.
The last term describes the cross-Kerr interaction.
By combining it with the first term as $(\hbar\omega_\text{H}-\chi\hat{b}^\dagger\hat{b})\hat{a}^\dagger\hat{a}$, the dependence of the HF mode resonance on the number of photons in the LF mode becomes apparent.
%

The cross-Kerr interaction manifests as photon-number splitting~\cite{Schuster2007} in the measured microwave reflection $S_{11}$ (Fig.~1D).
Distinct peaks correspond to the first transition frequency of the HF mode $|g,n\rangle\leftrightarrow|e,n\rangle$, with frequencies $\omega_\text{H}-n\chi/\hbar$ where $\chi/h =21$~MHz.
We label the eigenstates of the system $|j,n\rangle$, with $j=g,e,f, ...$ ($n=0,1,2, ...$ ) corresponding to excitations of the HF (LF) mode.
The amplitude of peak $n$ is proportional to
\begin{equation}
  P_n\kappa_\text{ext}/\kappa_n\ ,
  \label{eq:peak_amplitude}
\end{equation}
where $P_n$ is the occupation of photon-number level $|n\rangle$ in the LF mode and $\kappa_\text{ext}/\kappa_n$ is the ratio of external coupling $\kappa_\text{ext}/2\pi=1.6\cdot 10^6s^{-1}$ to the total line-width $\kappa_n$ of peak $n$.
From the Bose-Einstein distribution of peak heights $P_n$, we extract the average photon occupation $n_\text{th} = 1.6$ corresponding to a mode temperature of $17$~mK.
The resolution of individual photon peaks is due to the condition $\kappa_n\ll\chi/\hbar$.
The peak line-widths increase with $n$ following $\kappa_n = \kappa(1+4 n_\text{th}^{(H)})+2\gamma(n+(1+2n)n_\text{th})$, where $\kappa/2\pi=3.7\cdot 10^6s^{-1}$ is the dissipation rate of the HF mode, $n_\text{th}^{(H)}\simeq0.09$ its thermal occupation (see Fig.~S10), and $\gamma/2\pi=23\cdot 10^3s^{-1}$ is the dissipation rate of the LF mode (obtained through time-domain measurement Fig.~4).
The condition $\kappa_n\ll A_\text{H}/\hbar$ makes the HF mode an inductively-shunted transmon qubit~\cite{Koch2007}, making it possible to selectively drive the $|g,n\rangle\leftrightarrow|e,n\rangle$ and $|e,n\rangle\leftrightarrow|f,n\rangle$ transitions.
Despite its low dissipation rate $\gamma$, the LF mode has a line-width of a few~MHz (measured with two-tone spectroscopy, Fig.~S15) which originates in thermal processes such as $|g,n\rangle\rightarrow|e,n\rangle$ occurring at rates $\sim\kappa n_\text{th}^{(H)}$ larger than $\gamma$~\cite{SI}.
The LF mode line-width is then an order of magnitude larger than $A_\text{L}$, making it essentially a harmonic oscillator that we will refer to as the resonator.

The junction non-linearity enables transfer of population between states by coherently pumping the circuit at a frequency $\omega_\text{p}$.
%
%
The cosine potential of the junction imposes four-wave mixing selection rules, only allowing interactions that involve 4 photons.
One such interaction is
\begin{equation}
  \begin{split}
    \hat H_\text{int}=&-\hbar g\sqrt{n+1}|f,n\rangle\langle g,n+1|+h.c.\ ,\\
  \end{split}
  \label{eq:cooling_int}
\end{equation}
activated when driving at the energy difference between the two coupled states $\omega_\text{p}=2\omega_\text{H}-\omega_\text{L}-\left(2n\chi+A_\text{H}\right)/\hbar$.
This process, enabled by a pump photon, annihilates a photon in the resonator and creates two in the transmon. 
The number of photons involved in the interaction is four, making it an allowed four-wave mixing process.
The induced coupling rate is $g=A_\text{H}^{\frac{3}{4}}A_\text{L}^{\frac{1}{4}}\xi_\text{p}$, where $|\xi_\text{p}|^2$ is the amplitude of the coherent pump tone measured in number of photons~\cite{SI}.

We use this pump tone in combination with the large difference in mode relaxation rates to cool the megahertz resonator to its ground-state (Fig.~2A).
The pump drives transitions between $|g,1\rangle$ and $|f,0\rangle$ at a rate $g$.
The population of $|g,1\rangle$, transfered to $|f,0\rangle$, subsequently decays at a rate $2\kappa$ to the ground-state $|g,0\rangle$.
Cooling occurs when the thermalization rate of the resonator $n_\text{th}\gamma$ is slower than the rate $C\gamma$ at which excitations are transfered from $|g,1\rangle$ to $|g,0\rangle$, where $C = 2g^2/\kappa\gamma$ is the cooperativity (proportional to cooling-pump power~\cite{SI}).

%
%
For different cooling pump strengths, we measure $S_{11}$ (Fig.~2B).
The pump frequency is adapted at each power since the AC-stark effect increasingly shifts the qubit frequency as a function of power (see Fig.~S9).
The data is fitted to a sum of complex Lorentzians, with amplitudes given by Eq.~(\ref{eq:peak_amplitude}) and line-widths $\kappa_n$, from which $P_n$ is extracted.
Thermal effects lead to the ratio $P_{n+1}/P_n = n_\text{th}/(1+n_\text{th})$ between neighboring photon-number states for $n\ge 1$, and the cooling pump changes the ratio of occupation of the first two states
\begin{equation}
  \frac{P_1}{P_0} \simeq \frac{n_\text{th}}{1+n_\text{th}+C}\ .
\label{eq:main_P1_over_P0}
\end{equation}
The ground-state occupation hence increases with cooperativity and we attain a maximum $P_0=0.82$.
At higher cooperativity, $P_0$ diminishes due to the off-resonant driving of other four-wave mixing processes such as $|f,n+1\rangle\langle g,n|+h.c.$ which tend to raise the photon number of the resonator.
This effect is simulated using an adaptive rotating-wave approximation~\cite{baker2018adaptive} (Fig.~2C and S6).

Neighbouring four-wave mixing processes are measured by sweeping the pump frequency whilst monitoring the spectrum (Fig.~3A).
When cooling with a single pump they eventually limit performance, but can be resonantly driven to our advantage.
By driving multiple cooling interactions $|g,n\rangle\leftrightarrow|f,n-1\rangle$, less total pump power is required to reach a given ground-state occupation, hence minimizing off-resonant driving.
By maximizing the ground-state peak amplitude as a function of the power and frequency of four cooling tones, we achieve $P_0 = 0.90$ (Fig.~3B).
By combining cooling $|g,n\rangle\leftrightarrow|f,n-1\rangle$ and raising $|g,n\rangle\leftrightarrow|f,n+1\rangle$ tones (inset of Fig.~3C), we demonstrate stabilization of higher Fock states, non-Gaussian states commonly considered as non-classical phenomena~\cite{rips2012steady}.
The optimum frequencies for the raising and cooling tones adjacent to the stabilized state were detuned by a few~MHz from the transition frequency (see dashed lines in the inset of Fig.~3C), otherwise one pump tone would populate the $|f\rangle$ level, diminishing the effectiveness of the other.

%
Finally we investigate dynamics in a photon resolved manner (Fig.~4).
Whilst probing $S_{11}$ at a given frequency, we switch the cooling or single photon stabilization pumps on and off for intervals of $50\ \upmu$s.
We perform this for a sequence of probe frequencies, resulting in $S_{11}$ as a function of both frequency and time (see full spectrum in~\cite{SI}).
The spectrum is fitted at each time to extract $P_n$ as a function of time.
After reaching the steady state, the pumps are turned off and we observe the thermalization process which follows the semi-classical master equation
\begin{equation}
\begin{split}
  \dot P_n &= -n \gamma (n_\text{th}+1) P_n + n\gamma n_\text{th}P_{n-1}\\
  & -(n+1)P_n\gamma n_\text{th} + (n+1)P_{n+1}\gamma (n_\text{th}+1)\ .
\end{split}
\label{eq:main_rate_equation}
\end{equation}
Our cQED architecture enables the readout and manipulation of a radio-frequency resonator at the quantum level.
Utilizing the fast readout methods of cQED, single-shot readout or the tracking of quantum trajectories could enable even finer resolution of thermodynamic effects at the quantum scale.
Coupling many of these devices together could enable the exploration of many-body effects in Bose-Hubbard systems with dynamically tunable temperatures~\cite{rigol2008thermalization,sorg2014relaxation}.
This circuit architecture could also be used to interface circuit quantum electrodynamics with different physical systems in the MHz frequency range, such as spin systems~\cite{ares2016sensitive} or macroscopic mechanical oscillators\cite{Teufel2011}.
Finally, this circuit could enable sensing of radio frequency radiation with quantum resolution, a critical frequency range for a number of applications, from nuclear magnetic resonance imaging to radio astronomy.

\textbf{Acknowledgments: }
We acknowledge Ya. M. Blanter, S. M. Girvin, J. D. P. Machado for useful discussions.
\textbf{Funding: }
This work was supported by the European Research Council under the European Union’s H2020 program under grant agreements 681476 - QOM3D and 785219 - GrapheneCore2, by the Dutch Foundation for Scientific Research (NWO) through the Casimir Research School, and by the Army Research Office through Grant No.\ W911NF-15-1-0421.
\textbf{Author contributions: }
MFG and RV developed the theoretical description of the experiment. 
MFG designed the device. 
MFG fabricated the device with help from JD and MK. 
MFG, MK, CD, JD and MJ participated in the measurements.
MFG and CD analyzed the data.
BB provided the software and input for the adaptive rotating-wave-approximation simulation.
MFG wrote the manuscript with input from MK, CD and GAS.
All co-authors reviewed the manuscript and provided feedback.
GAS supervised the project.
\textbf{Competing interests: }
The authors declare that they have no competing interests.
\textbf{Data and materials availability: }
Raw data as well as all measurement, data-analysis and simulation code used in the generation of main and supplementary figures is available in Zenodo with the identifier 10.5281/zenodo.2551258

\bibliography{library}
\let\addcontentsline\oldaddcontentsline

\clearpage

\onecolumngrid
\begin{center}
{\huge \textbf{Supplementary information}}
\end{center}
\vspace{-20pt}
\twocolumngrid




\makeatletter
   \renewcommand\l@section{\@dottedtocline{2}{1.5em}{1em}}
   \renewcommand\l@subsection{\@dottedtocline{2}{3.5em}{1em}}
   \renewcommand\l@subsubsection{\@dottedtocline{2}{5.5em}{1em}}
\makeatother


\renewcommand{\thesection}{\arabic{section}} 

\onecolumngrid
\tableofcontents

\let\oldaddcontentsline\addcontentsline
\renewcommand{\addcontentsline}[3]{}
\listoffigures
\listoftables
\let\addcontentsline\oldaddcontentsline


\renewcommand{\theequation}{S\arabic{equation}}
\renewcommand{\thefigure}{S\arabic{figure}}
\renewcommand{\thetable}{S\arabic{table}}
\setcounter{figure}{0}
\setcounter{equation}{0}

\newcolumntype{C}[1]{>{\centering\arraybackslash}p{#1}}
\newcolumntype{L}[1]{>{\raggedright\arraybackslash}p{#1}}


\section{Fabrication}
\begin{figure*}[ht!]
\includegraphics[width=1\columnwidth]{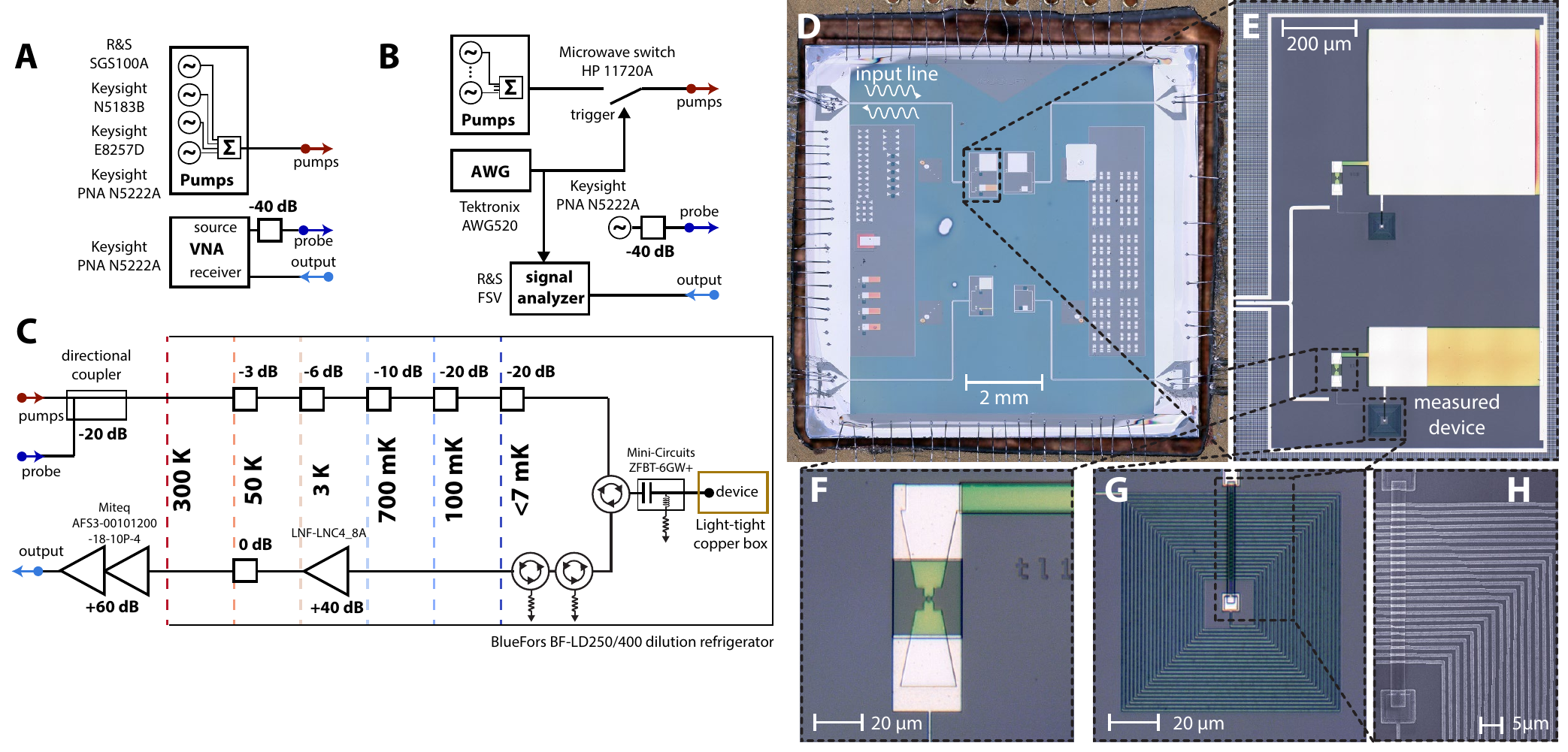}
\caption[Experimental setup and device]{\textbf{Experimental setup and device. A: }
room temperature setup for spectroscopy experiments, \textbf{B: } room temperature setup for time-domain experiments.
These setups are connected to the fixed setup shown in \textbf{C}.
\textbf{C: }Cryogenic setup.
\textbf{D: }Optical image of the chip, wire-bonded to a surrounding printed circuit board (PCB).
The PCB is mounted in a copper box which is cooled below $7$~mK (\textit{i.e.} under the range of our fridge thermometry) in our dilution refrigerator.
\textbf{E: }Optical image of the two circuits connected to the measured feedline.
Due a small cross-Kerr to line-width ratio, photon-number splitting was not achieved in the top device, where the low (high) mode was designed to resonate at $\sim50$~MHz ($\sim7.2$ GHz).
\textbf{F: }Optical image of the SQUID, under a protective a-Si:H layer to avoid damage from Ar milling in the last step.
\textbf{G,H: }Optical and SEM image of the 23-turn spiral inductor which has a $1.5\ \upmu$m pitch and a $500\ $nm wire width.
}
\label{fig:S_setup}
\end{figure*}

The circuit is fabricated on a high resistivity silicon substrate using aluminum as a superconductor.
Using shadow evaporation, we first pattern Al/AlOx/Al Josephson junctions, the bottom plate of the capacitor and an underpass (a line connecting the center of the spiral inductor to the capacitor).
We then deposit $260$ nm of hydrogenated amorphous silicon (a-Si:H) as a dielectric layer, motivated by its expected low dielectric loss~\cite{o2008microwave}, using PECVD (plasma enhanced chemical vapor deposition)
The a-Si:H is patterned to form a dielectric layer for the parallel plate capacitor, a bridge over the spirals underpass, and a protection layer above the junctions.
Finally we sputter-deposit and pattern aluminum to form the rest of the circuitry, after an argon-milling step to ensure a galvanic connection to the first aluminum layer.
The resulting circuit is shown in detail in Fig.~\ref{fig:S_setup}.

\section{Experimental setup}

\section{Data filtering}
In Figs~1A, 3A, \ref{fig:S_caveats}B, \ref{fig:S_temperature}B,C, \ref{fig:S_all_transitions}, \ref{fig:S_time_domain}A,B,C we applied a Gaussian filter with a standard deviation of one increment in the x-axis (and y-axis when applicable).
The filtering was used in the construction of the figure for clarity.
No filtering was applied before fitting the data.


\section{Theory}
\twocolumngrid
\setlength{\parskip}{1em}

\begin{figure*}[ht!]
\includegraphics[width=0.7\textwidth]{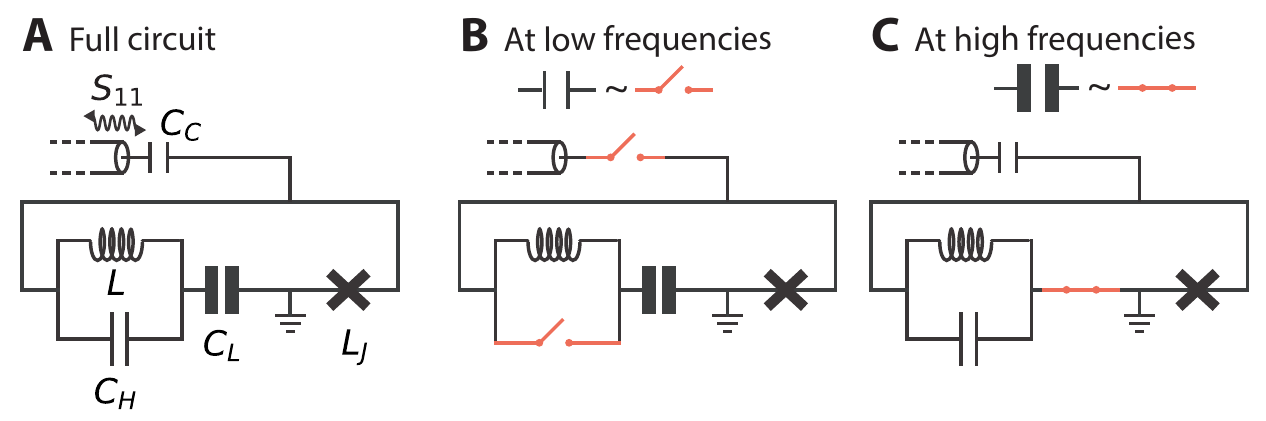}
\caption[Detailed circuit diagram]{The circuit studied in this work (A) and approximate circuits for the low-frequency (B) and high-frequency regime (C).
}
\label{fig:S_circuit}
\end{figure*}

\subsection{Circuit Hamiltonian}
\label{sec:black_box}

\begin{figure*}[ht!]
\includegraphics[width=0.9\textwidth]{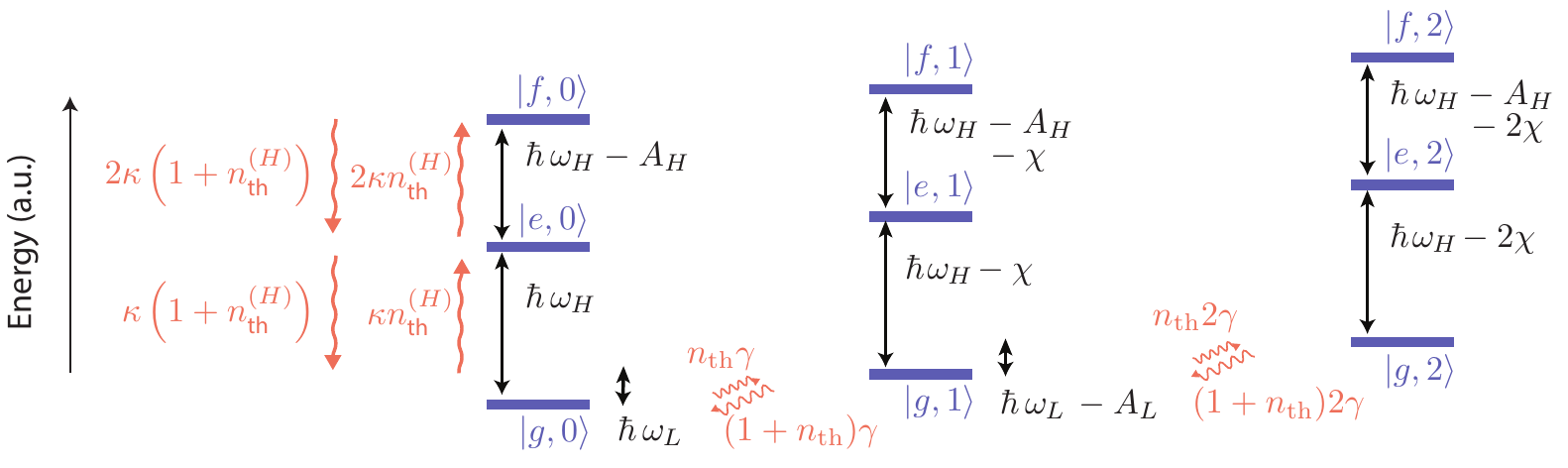}
\caption[Detailed energy diagram]{\textbf{Detailed energy diagram of the system}. 
We depict the first three levels of both high and low mode along with their dissipation and thermalization rates.
Transition energies are written with $\hbar=1$.
}
\label{fig:energy_diagram}
\end{figure*}

In this section, we derive the Hamiltonian for the circuit shown in Fig.~\ref{fig:S_circuit}A using the black-box quantization method~\cite{nigg_black-box_2012}.
This method allows the systematic derivation of the resonance frequency $\bar\omega_m$ and anharmonicity $A_m$ of the different modes $m$ of a circuit from the admittance $Y(\omega) = 1/Z(\omega)$ across the Josephson junction if we replace the latter by a linear inductor $L_\text{J} = \hbar^2/4 e^2 E_\text{J}$.
The resonance frequencies $\bar\omega_m$ are the zeros of the admittance $Y(\bar\omega_m) = 0$, and the anharmonicities are given by 
\begin{equation}
	A_m = -\frac{2e^2}{L_\text{J}\bar\omega_m^2(\text{Im}Y'(\bar\omega_m))^2}\ .
	\label{eq:anharmonicity}
\end{equation}
The idea is to quantify through $A_m$ the amount of current traversing the Josephson junction for an excitation in mode $m$.
The Hamiltonian of the circuit is then
\begin{equation}
	\begin{split}
	\hat{H} &= \sum_m\hbar\bar\omega_m\hat{a}_m^\dagger\hat{a}_m + \underbrace{E_\text{J}[1-\cos{\hat{\varphi}}]-E_\text{J}\frac{\hat{\varphi}^2}{2}}_\text{junction non-linearity}\ ,\\
	\text{where }\hat{\varphi} &= \sum_m\left(2 A_m/E_\text{J}\right)^{1/4}(\hat{a}_m^\dagger+\hat{a}_m)\ .
	\label{eq:Hamiltonian_8th_order}
\end{split}
\end{equation}
In the circuit of Fig.~\ref{fig:S_circuit}A , there are two modes, a high-frequency one and a low-frequency one.
By comparing to a black-box quantization of the full circuit, we find that taking the approximation of $C_H\simeq0$, $C_c\simeq0$ for the low-frequency mode and $C_L\simeq\infty$ for the high-frequency mode results in corrections of only $0.2$, $1.2$, $0.3$ and $2.1$ \% in the value of $\omega_L$, $\omega_H$, $A_L$ and $A_H$ respectively.
It is therefore a good approximation, which has the additional advantage of producing simple analytical equations for the frequencies and anharmonicities of the circuit.
Starting with the low-frequency mode shown in Fig.~\ref{fig:S_circuit}B, we find the (imaginary part of the) admittance across the linearized junction to be
\begin{equation}
	\text{Im}Y(\omega) = \frac{1}{\omega L_J}\frac{\left(\frac{\omega}{\omega_L}\right)^2-1}{1-\left(\omega\sqrt{LC_L}\right)^2}\ ,
\end{equation}
yielding the resonance frequency
\begin{equation}
	\omega_L = \frac{1}{\sqrt{(L+L_J)C_L}}\ .
	\label{eq:wl}
\end{equation}
Taking the derivative of the imaginary part of the admittance at $\omega = \omega_L$ yields:
\begin{equation}
	\text{Im}\frac{\partial Y}{\partial \omega}(\omega_L) = 2C_L\left(\frac{L+L_J}{L_J}\right)^2
\end{equation}
Substituting this into Eq.~(\ref{eq:anharmonicity}) yields 
\begin{equation}
	A_L = -\frac{e^2}{2C_L}\left(\frac{L_J}{L+L_J}\right)^3\ .
	\label{eq:Al}
\end{equation}
Turning to the high-frequency mode shown in Fig.~\ref{fig:S_circuit}C, we find the (imaginary part of the) admittance across the linearized junction to be 
\begin{equation}
	\text{Im}Y(\omega) = C_H\omega\left(1-\frac{\omega_H^2}{\omega^2}\right)\ ,
\end{equation}
yielding the resonance frequency
\begin{equation}
	\omega_H = \sqrt{\frac{L+L_J}{LL_JC_H}}\ .
	\label{eq:wh}
\end{equation}
Taking the derivative of the imaginary part of the admittance at $\omega = \omega_H$ yields:
\begin{equation}
	\text{Im}\frac{\partial Y}{\partial \omega}(\omega_H) = 2C_H
\end{equation}
Substituting this into Eq.~(\ref{eq:anharmonicity}) yields 
\begin{equation}
	A_H = -\frac{e^2}{2C_H}\left(\frac{L}{L+L_J}\right)\ .
	\label{eq:Ah}
\end{equation}
A Taylor expansion of the junctions cosine potential is justified if the anharmonicities are weak and only a few photons populate the circuit.
Whilst numerical calculations in this work consider the 8-th order expansion, much understanding can be gleaned by stopping the expansion at the fourth-order
\begin{equation}
\begin{split}
		\hat{H}_{4,\text{diag}} =& \hbar\omega_\text{H}\hat{a}^\dagger\hat{a} -\frac{A_\text{H}}{2}\hat{a}^\dagger\hat{a}^\dagger\hat{a}\hat{a}\\
		&+ \hbar\omega_\text{L}\hat{b}^\dagger\hat{b} -\frac{A_\text{L}}{2}\hat{b}^\dagger\hat{b}^\dagger\hat{b}\hat{b}\\
		&-\chi\hat{a}^\dagger\hat{a}\hat{b}^\dagger\hat{b}\ ,\\
		\label{eq:H4_diag}
\end{split}
\end{equation}
where $\chi$ is the cross-Kerr coupling: the amount by which the high-mode transition shifts as a result of adding an excitation in the low mode and vice versa.
We defined the first transition frequencies of both modes
\begin{equation}
\begin{split}
	\hbar\omega_\text{H}=\hbar\bar\omega_\text{H}-A_\text{H}-\frac{\chi}{2}\ ,&\\
	\hbar\omega_\text{L}=\hbar\bar\omega_\text{L}-A_\text{L}-\frac{\chi}{2}\ .&\\
\end{split}
\end{equation}
In Eq.~(\ref{eq:H4_diag}), we have neglected terms in the expansion which are off-diagonal in the Fock basis and do not modify the eigenergies to leading order perturbation theory.
The eigenfrequencies of the system are summarized in the energy diagram of Fig.~\ref{fig:energy_diagram}
\subsection{Comparison to the typical circuit QED architecture}

We now compare our circuit architecture with a more conventional circuit QED coupling scheme~\cite{Koch2007}, where the transmon qubit with a frequency $\omega_\text{H}$ couples capacitively at a rate $G$ to an $LC$-oscillator ($\omega_\text{L}$).
In this architecture, the cross-Kerr coupling would be $4A_\text{H} (\bar g\omega_\text{L}/\omega_\text{H}^2)^2$, to first order in $\bar G$ and $A_\text{H}$~\cite{gely2017nature}.
If we would want a cross-Kerr coupling to exceed $\kappa$, for the large frequency difference $\omega_\text{L}\ll\omega_\text{H}$ demonstrated in this work, the couplings $G$ would have to be very large.
As is well known from ultra-strong coupling circuit QED architectures, this translates to both high impedance resonators~\cite{bosman2017approaching} and large coupling capacitors~\cite{bosman_multi-mode_2017}.
These elements are all present in this circuit if we consider $C_L$ as a coupling capacitor between the high impedance $L_\text{H}C_\text{H}$-oscillator and the qubit constituted of the junction and the junctions capacitance (that we have neglected in the previous Hamiltonian derivation).
However, the basis of normal modes presented in the previous section present a much more convenient framework to understand the system.

\subsection{Translating the measured $S_{11}$ to a quantum operator}

We now introduce a driving term in the Hamiltonian and consider losses to both the environment and the measurement port.
Following input-output theory~\cite{vool2017introduction,clerk2010introduction}, the quantum Langevin equation for $\hat{a}(t)$ is
\begin{equation}
	\frac{\text{d}}{\text{d}t}\hat{a}(t) = \frac{i}{\hbar}[\hat{H}_\text{undr},\hat{a}(t)] - \frac{\kappa}{2}\hat{a}(t)+\sqrt{\kappa_\text{ext}}\tilde{a}_\text{in}(t)\ .
\end{equation}
Where the undriven Hamiltonian $\hat{H}_\text{undr}$ corresponds to that of Eq.~(\ref{eq:Hamiltonian_8th_order}), where the degree of expansion of the non-linearity is yet unspecified.
The microwave reflection measured in spectroscopy (here in the time-domain) is given by
\begin{equation}
	S_\text{11}(t) = \frac{\tilde{a}_\text{out}(t)}{\tilde{a}_\text{in}(t)} = 1-\sqrt{\kappa_\text{ext}}\frac{\hat{a}(t)}{\tilde{a}_\text{in}(t)}\ ,
	\label{eq:in_out}
\end{equation}
where $\tilde{a}_\text{in}(t)$ ($\tilde{a}_\text{out}(t)$) is the incoming (outgoing) field amplitude, $\kappa_\text{ext}$ ($\kappa$) is the external (total) coupling rate of the high-frequency mode.
The coupling of the low mode to the feedline $\gamma_\text{ext}/2\pi=2s^{-1}$ is much smaller than coupling of the high mode to the feedline $\kappa_\text{ext}/2\pi=1.63\cdot10^6s^{-1}$, we therefore assume that a drive tone only affects the high-frequency mode.
For a coherent drive, characterized by a drive frequency $\omega_\text{d}$ and an incoming power $P_\text{in}$ (equal to the average power $\langle P(t) \rangle$ of the oscillating input signal), the wave amplitude is
\begin{equation}
	\tilde{a}_\text{in}(t)=\sqrt{\frac{P_\text{in}}{\hbar\omega_\text{d}}}  e^{-i\omega_\text{d} t}\ ,
\end{equation}
and the drive term can be incorporated in the Hamiltonian of the system
\begin{align}
\begin{split}
	\frac{\text{d}}{\text{d}t}\hat{a}(t) &= \frac{i}{\hbar}[\hat{H}_\text{undr}+\hat{H}_\text{drive},\hat{a}(t)] - \frac{\kappa_\text{ext}}{2}\hat{a}(t)\ ,\\
	\text{where }\hat{H}_\text{drive} &= i\hbar\epsilon_\text{d}\left(e^{-i\omega_\text{d} t}\hat{a}^\dagger(t) - e^{i\omega_\text{d} t}\hat{a}(t)\right)\ ,\\
	\epsilon_\text{d} &= \sqrt{\frac{\kappa_\text{ext}P_{in}}{\hbar\omega_\text{d}}}\ .
\end{split}
\end{align}
%
Additionally, we also remove the time-dependence in the drive Hamiltonian by moving to a frame rotating at $\omega_\text{d}$ with the unitary transformation $U_\text{probe} = e^{i\omega_\text{d} t \hat{a}^\dagger\hat{a}}$, 
\begin{equation}
	\frac{\text{d}}{\text{d}t}\hat{a} = \frac{i}{\hbar}[U_\text{probe}^\dagger\hat{H}_\text{undr}U_\text{probe}+\tilde{H}_\text{drive},\hat{a}] - \frac{\kappa_\text{ext}}{2}\hat{a}\ ,
	\label{eq:langevin}
\end{equation}
where $\hat a e^{i\omega_d t} = \hat a(t)$ and 
\begin{equation}
	\tilde{H}_\text{drive} =  -\hbar\omega_\text{d} \hat{a}^\dagger  \hat{a}\\
	+i\hbar\epsilon_\text{d}\left(\hat{a}^\dagger -\hat{a}\right)\ .
	\label{eq:drive_hamiltonian}
\end{equation}
In this rotating frame, the reflection coefficient becomes
\begin{equation}
	\hat{S}_\text{11}(\omega_\text{d}) = 1-\frac{\kappa_\text{ext}}{\epsilon_\text{d}}\hat{a}\ ,
	\label{eq:in_out_freq}
\end{equation}
of which we measure the expectation value when probing the system.
From now on, and in the main text we use the shorthand $S_{11}(\omega_\text{d}) = \langle\hat{S}_\text{11}(\omega_\text{d})\rangle$.
Note that by casting the quantum Langevin Eq.~(\ref{eq:langevin}) in the form
\begin{equation}
\begin{split}
	\frac{\text{d}}{\text{d}t}\hat{a} =& \frac{i}{\hbar}[U_\text{probe}^\dagger\hat{H}_\text{undr}U_\text{probe}+\hat{H}_\text{drive},\hat{a}]\\ 
	+&\left(L^{\dagger}\hat{a}L - \frac{1}{2}\left(\hat{a}L^{\dagger}L + L^{\dagger}L\hat{a}\right)\right)\ ,\\
	\text{where }L=&\kappa_\text{ext} \hat{a}\ ,
\end{split}
\end{equation}
it can be readily transformed to a Lindblad equation
\begin{equation}
\begin{split}
	\frac{\text{d}}{\text{d}t}\rho =& -\frac{i}{\hbar}[U_\text{probe}^\dagger\hat{H}_\text{undr}U_\text{probe}+\hat{H}_\text{drive},\rho] \\
	+& \left(L \rho L^{\dagger}-\frac{1}{2}\left(\rho L^{\dagger}L + L^{\dagger}L\rho\right)\right)\ \\
	=& -\frac{i}{\hbar}[U_\text{probe}^\dagger\hat{H}_\text{undr}U_\text{probe}+\hat{H}_\text{drive},\rho] + \kappa_\text{ext}\mathcal{L}[\hat{a}]\ ,
	\label{eq:probed_lindbald_early}
\end{split}
\end{equation}
better suited to numerical calculations using QuTiP~\cite{johansson2012qutip}.

\begin{figure*}[ht!]
\includegraphics[width=0.53\textwidth]{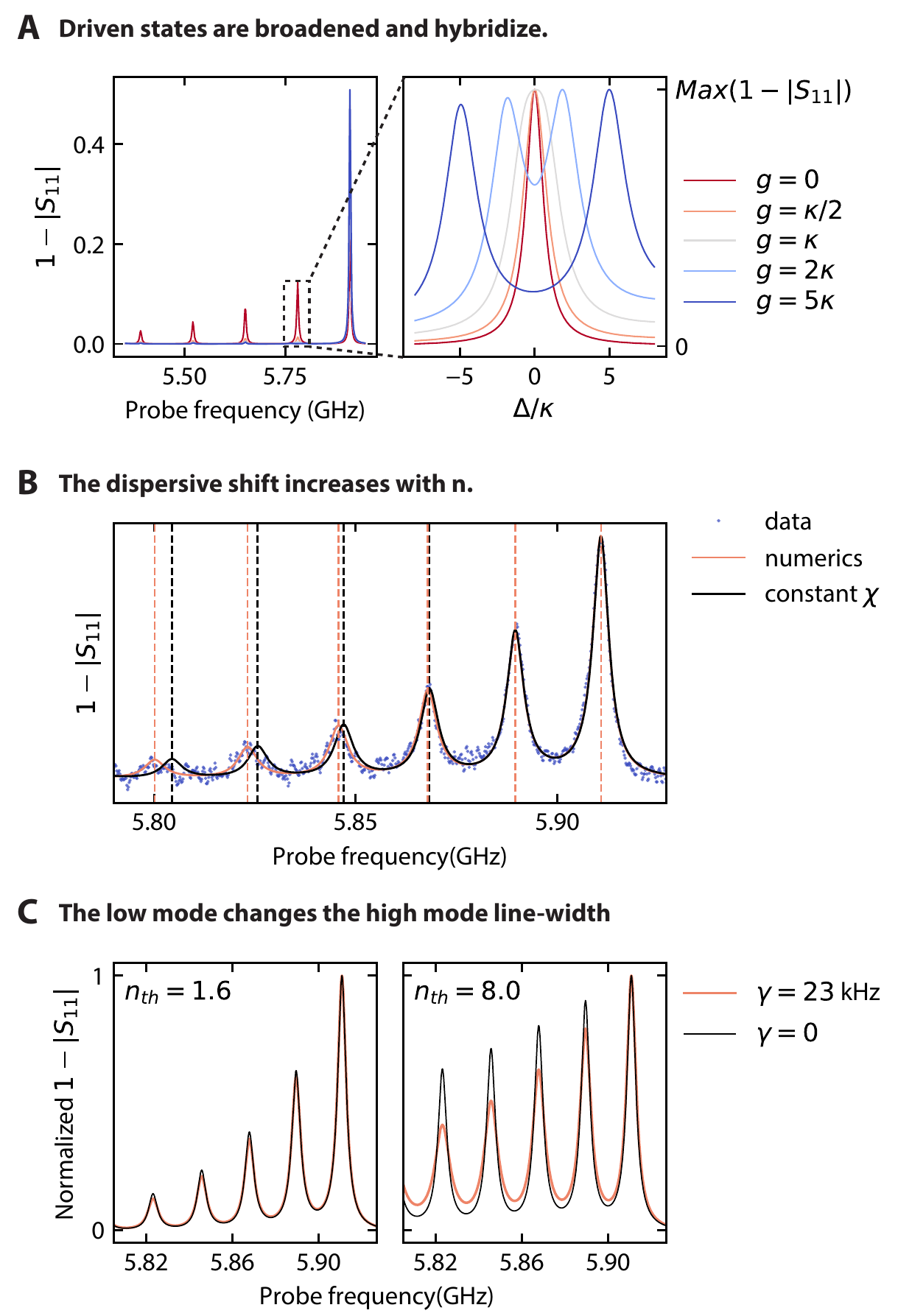}
\caption[Possible caveats in fitting a sum of Lorentzians to $S_{11}$]{\textbf{Possible caveats in fitting a sum of Lorentzians to $S_{11}$}.
\\\hspace{\textwidth}
\textbf{A: driven states are broadened then hybridize.}
As we increase the coupling $g$ induced by a cooling pump resonant with $\ket{g,1}\leftrightarrow\ket{f,0}$,the low mode is cooled as shown in the left panel.
In the right panel, we zoom in to the normalized $n=1$ peak.
As a consequence of the coupling between levels $\ket{g,1}$ and $\ket{f,0}$, this peak first broadens then splits into two distinct peaks,
The slight asymmetry arises from the tail of the $n=0$ peak.
We used the device parameters with an increased $\chi= h\times130$~MHz and $A_\text{H}= h\times600$~MHz in order to minimize the visibility of the tail of the other peaks.
\\\hspace{\textwidth}
\textbf{B: the dispersive shift increases with $n$.}
The Hamiltonian presented in Eq.~(2), which only considers the diagonal contributions of the quartic term of the JJ non-linearity, results in a constant shift of the high-mode frequency $\omega_H-n\chi/\hbar$.
As shown in black, overlaid on the blue dots of the same data as in panel A, this results in a slight misalignment of the peaks.
By diagonalizing the Hamiltonian of Eq.~(\ref{eq:Hamiltonian_8th_order}), with the JJ non-linearity Taylor expanded to the 8-th order, we achieve a more realistic prediction of the system frequencies, and find that the shift increases with the number of photons in the low mode, as shown with red lines.
\\\hspace{\textwidth}
\textbf{C: $\gamma$ and $n_\text{th}$ modify the high-mode line-width $\kappa_{j,n}$.}
As shown in Eq.~(\ref{eq:sum_of_lorentzians}), the high-mode line-width not only depends on high-mode dissipation rate $\kappa$, but also on the dissipation $\gamma$ and thermal occupation $n_\text{th}$ of the low mode.
As $\gamma\ll\kappa$, this effect is subtle for low thermal occupations, but if neglected, can lead to an underestimation of the low mode occupation at higher temperatures.
}
\label{fig:S_caveats}
\end{figure*}

\subsection{Derivation of $S_{11}[\omega]$: probing the system}

In this section we derive the spectrum of the high mode for arbitrary states of the low mode.
We append the Lindblad equation of Eq.~(\ref{eq:probed_lindbald_early}) to take into account additional interactions of the system with the environment.
Internal dissipation of the high mode $\kappa_\text{int}$, is added to the external dissipation rate to constitute its total dissipation rate $\kappa = \kappa_\text{int}+\kappa_\text{ext}$.
The low mode is attributed a dissipation rate $\gamma$.
The average thermal occupation of the two modes are denoted by $n_\text{th}^{(H)}$ and $n_\text{th}$ for the high and low mode respectively.
We can estimate the response function $S_{11}(\omega_\text{d})$ analytically using the Hamiltonian of Eq.~(\ref{eq:H4_diag}).
The unitary $U_\text{probe}$ leaves this Hamiltonian unchanged and the complete Lindblad equation is then
\begin{equation}
\begin{split}
	\frac{\text{d}}{\text{d}t} \rho = &-\frac{i}{\hbar}[H_\text{4,diag}+\hat{H}_\text{drive},\rho] \\
	&+\kappa (n_\text{th}^{(H)}+1) \mathcal{L}[\hat{a}]
	+\kappa n_\text{th}^{(H)} \mathcal{L}[\hat{a}^\dagger]\\
	&+\gamma (n_\text{th}+1) \mathcal{L}[\hat{b}]
	+\gamma n_\text{th} \mathcal{L}[\hat{b}^\dagger]\ .
	\label{eq:probed_lindbald}
\end{split}
\end{equation}
In the un-driven case $\epsilon_\text{d}=0$, we assume the steady-state solution to be a diagonal density matrix $\rho_\text{ss}$ as a consequence of thermal effects
\begin{equation}
	\rho_\text{ss} = 
\begin{bmatrix}
    P_g & 0&0 &    \\
    0 & P_e & 0 &   \\
    0 & 0 & P_f &   \\
     &  &  & \ddots \\
\end{bmatrix}_H
\otimes
\begin{bmatrix}
    P_0 & 0&0 &    \\
    0 & P_1 & 0 &   \\
    0 & 0 & P_2 &   \\
     &  &  & \ddots \\
\end{bmatrix}_L\ ,
\label{eq:thermal_steady_state}
\end{equation}
where $P_g,P_e,P_f$ ($P_0,P_1,P_2$) corresponds to the occupation of high (low) mode levels.
Note that when we pump the system, effectively coupling levels of the high and low mode, this approximation breaks down and that particular limit is discussed below.
We now look for a perturbative correction to this matrix at a small driving rate $\epsilon_\text{d}$
\begin{equation}
	\rho = \rho_\text{ss} + \epsilon_\text{d} \rho_\text{pert} \ ,
\end{equation}
where $\rho_\text{pert}$ has the unit time. 
The objective is to determine the expectation value of the reflection coefficient 
\begin{equation}
	S_\text{11}(\omega_\text{d}) = \text{Tr}\left[\rho \left(1-\frac{\kappa_\text{ext}\hat{a}}{\epsilon_\text{d}}\right)\right] \ .
\end{equation}
We substitute the perturbative expansion of $\rho$ into Eq.~(\ref{eq:probed_lindbald}) and keep only terms to first order in $\epsilon_\text{d}$.
This equation is solved analytically in reduced Hilbert-space sizes using the software Wolfram Mathematica.
The largest Hilbert-space sizes for which Mathematica could provide an analytical solution in a reasonable amount of time were: (4,0), (3,2), (2,5) where the first (second) number designates the number of levels included in the high (low) mode.
We extrapolate the obtained results to construct the reflection coefficient
\begin{equation}
\begin{split}
	S_\text{11}(\omega_\text{d}) &= 1-(P_g-P_e)\sum_n P_n\frac{\kappa_\text{ext}}{i\Delta_{g,n}+\kappa_{g,n}}\\
	&\ \ \ \ \ -(P_e-P_f)\sum_n P_n\frac{2\kappa_\text{ext}}{i\Delta_{e,n}+\kappa_{e,n}}\ ,\\
	\text{where }\kappa_{g,n} &= \kappa(1+4 n_\text{th}^{(H)})+2\gamma(n+(1+2n)n_\text{th})\ ,\\
	\kappa_{e,n} &= \kappa(3+8 n_\text{th}^{(H)})+2\gamma(n+(1+2n)n_\text{th})\ .\\
\end{split}
\label{eq:sum_of_lorentzians}
\end{equation}
which corresponds to a sum of Lorentzian functions, each associated to high-mode level $i$ and a low-mode level $n$, with line-width $\kappa_{i,n}$ centered around $\Delta_{i,n} = 0$, where
\begin{equation}
 	\begin{split}
 		&\Delta_{g,n} = \omega_\text{H}-n\chi/\hbar-\omega_\text{d}\ ,\\
		&\Delta_{e,n} = \omega_\text{H}-(n\chi-A_\text{H})/\hbar-\omega_\text{d}\ .
 	\end{split}
 	\label{eq:deltas}
 \end{equation} 
Note that in the main text we use the notation $\kappa_{g,n}=\kappa_n$.

%
By numerically computing $S_{11}$ as described in Sec.~\ref{sec:numerics_S11}, we find that the expression for the line-widths $\kappa_{i,n}$ remains accurate, whilst the center of the Lorentzians $\Delta_{i,n}$ will slightly shift from Eq.~(\ref{eq:deltas}), as shown in Fig.~\ref{fig:S_caveats}B.
When fitting data, we hence use the Eqs.~(\ref{eq:sum_of_lorentzians}) whilst fixing $\Delta_{i,n}$ with a diagonalization of the Hamiltonian Eq.~(\ref{eq:Hamiltonian_8th_order}) Taylor expanded to the 8-th order.
In Fig.~\ref{fig:S_temperature}(C), we show that Eq.~(\ref{eq:sum_of_lorentzians}) is in excellent agreement with both data and numerics.

\subsubsection{The impact of a pump tone on $S_{11}(\omega)$}
Pump tones can invalidate Eq.~(\ref{eq:sum_of_lorentzians}) in different ways.
As an example let us take the cooling scheme where a pump tone couples the levels $\ket{g,1}$ and $\ket{f,0}$ at a rate $g$.
This is simulated by numerically finding the steady state of the Hamiltonian
\begin{equation}
\begin{split}
		\hat{H} =& \hbar\Delta\hat{a}^\dagger\hat{a} -\frac{A_\text{H}}{2}\hat{a}^\dagger\hat{a}^\dagger\hat{a}\hat{a}\\
		&+ (2\hbar\Delta-A_\text{H}-A_\text{L})\hat{b}^\dagger\hat{b} -\frac{A_\text{L}}{2}\hat{b}^\dagger\hat{b}^\dagger\hat{b}\hat{b}\\
		&-\chi\hat{a}^\dagger\hat{a}\hat{b}^\dagger\hat{b}\\
		&+g(\ket{g,1}\bra{f,0}+\ket{f,0}\bra{g,1})\\
		&+i\hbar\epsilon_\text{d}(\hat{a}^\dagger-\hat a) ,\\
\end{split}
\end{equation}
written in a frame rotating at the probe frequency $\Delta = \omega_H-\omega_d$ and where the levels $\ket{g,1}$ and $\ket{f,0}$ are made resonant.
As shown in Fig.~\ref{fig:S_caveats}A, a peak corresponding to a transition to or from a level which is being pumped will be broadened in line-width and eventually will split into two peaks with increasing $g$.
This is not an issue in the cooling scheme since we do not use the driven $n=1$ peak to extract Fock-state fidelity, only the $n=0$ peak.

We do however off-resonantly pump $\ket{g,0}\leftrightarrow\ket{f,1}$ for example, along with many other transitions involving either state $\ket{g,0}$ or $\ket{e,0}$.
Off-resonant pumping should also lead to line-width broadening, this time of a peak used in extracting a Fock state fidelity.
To mitigate this issue we extract $P_n$ -- when stabilizing the $n$-th Fock state -- by using a fixed line-width $\kappa_n$ defined in Eq.~(\ref{eq:sum_of_lorentzians}).
This means that we always give a lower bound to $P_n$.
By comparing the pumped and un-pumped line-width of $n=0$ peak (see Fig.~\ref{fig:S_cooling}(B)), we notice no change in line-width with increasing pump power, indicating that our underestimation is certainly not drastic.
Finally, pump tones could drive the steady-state away from our assumption of a purely diagonal density matrix Eq.~(\ref{eq:thermal_steady_state}).
However we find that in the cooling experiment of Fig.~2, the adaptive rotating-wave simulation suggests that at maximum $P_0$, all off-diagonal terms of the density matrix are below $2.3\times10^{-3}$. 
This issue can safely be disregarded.

\subsection{Four wave mixing}
\subsubsection{Analytical derivation of the pump-induced coupling rates}

In this section we will consider the probe tone to be very weak and hence negligible.
Following Refs.~\cite{Leghtas2015}, we add a pump tone driving the high mode with frequency $\omega_\text{p}$ and strength $\epsilon_\text{p}$ to the system Hamiltonian
\begin{equation}
\begin{split}
	\hat{H}_\text{4,dr} &= \hbar\bar\omega_\text{H}\hat{a}^\dagger\hat{a}+\hbar\bar\omega_\text{L}\hat{b}^\dagger\hat{b} +E_\text{J}[1-\cos{\hat{\varphi}}]-\frac{E_\text{J}}{2}\hat{\varphi}^2 \\
	&+ \hbar \left(\epsilon_\text{p}e^{-i\omega_\text{p} t}+\epsilon_\text{p}^*e^{i\omega_\text{p} t}\right)\left(\hat{a}^\dagger +\hat{a}\right)\ ,\\
	\text{where }\hat{\varphi} &= \left(2 A_\text{H}/E_\text{J}\right)^{1/4}(\hat{a}^\dagger+\hat{a})+\left(2 A_\text{L}/E_\text{J}\right)^{1/4}(\hat{b}^\dagger+\hat{b})\ .
\end{split}
\end{equation}
We move to the displaced frame of the pump through the unitary transformation
\begin{equation}
	U_\text{pump} = e^{-\tilde{\xi}_\text{p}\hat{a}^\dagger+\tilde{\xi}_\text{p}^*\hat{a}}\ ,
\end{equation}
Where $\tilde{\xi}_\text{p}$ is defined by the differential equation
\begin{equation}
	\frac{d\tilde\xi_\text{p}}{dt}=-i\bar\omega_\text{H}\tilde\xi_\text{p}-i\left(\epsilon_\text{p}e^{-i\omega_\text{p} t}+\epsilon_\text{p}^*e^{i\omega_\text{p} t}\right)-\frac{\kappa}{2}\tilde\xi_\text{p}\ .
\end{equation}
For $t\gg1/\kappa$, and for far detuned drives $|\omega_\text{H}-\omega_\text{p}|\gg\kappa$, this equation is solved by
\begin{equation}
	\begin{split}
		\tilde\xi_\text{p}&\simeq{\epsilon_\text{p}e^{-i\omega_\text{p} t}}\left(\frac{1}{\omega_\text{p}-\bar\omega_\text{H}} +\frac{1}{\omega_\text{p}+\bar\omega_\text{H}} \right)\ .
	\end{split}
\end{equation}
In this frame, the Hamiltonian becomes
\begin{equation}
\begin{split}
	\hat{H}_\text{4,dr} =& \hbar\bar\omega_\text{H}\hat{a}^\dagger\hat{a}+\hbar\bar\omega_\text{L}\hat{b}^\dagger\hat{b} +E_\text{J}[1-\cos{\tilde{\varphi}}]-\frac{E_\text{J}}{2}\tilde{\varphi}^2 \\
	\text{where }\tilde{\varphi} =& \left(2 A_\text{H}/E_\text{J}\right)^{1/4}(\hat{a}^\dagger+\hat{a})+\left(2 A_\text{L}/E_\text{J}\right)^{1/4}(\hat{b}^\dagger+\hat{b})\\
	& +\left(2 A_\text{H}/E_\text{J}\right)^{1/4}(\tilde\xi_\text{p}^*+\tilde\xi_\text{p})
\end{split}
\end{equation}
We now Taylor expand the cosine non-linearity to fourth-order, neglecting terms which are off-diagonal in the Fock basis except when they depend on $\tilde\xi_\text{p}$ .
The latter can be made relevant depending on our choice of $\omega_\text{p}$.
\begin{equation}
\begin{split}
		\hat{H}_{4,\text{pumped}} &= \hat{H}_{4,\text{diag}} + \hat{H}_\text{p}\ ,\\
\end{split}
\label{eq:hamiltonian_pumped}
\end{equation}
Where $\hat{H}_{4,\text{diag}}$ was given in Eq.~(\ref{eq:H4_diag}).
The terms dependent on the pump power and frequency are assembled in the term $\hat{H}_\text{p}$ and written in Table \ref{tab:pump_terms}, along with the approximate pumping frequency $\omega_\text{p}$ necessary to eliminate their time-dependence.
As shown in the next paragraph, this occurs when the pump frequency matches the transition frequency between the two states coupled by the interaction term.

\begin{table}[t]
\centering
\caption[Four-wave mixing terms]{\textbf{Four-wave mixing terms}
Only half of terms are shown, the other half can be obtained by taking the hermitian conjugate of all these terms.
Terms become approximately time-independent around the frequency $\omega_\text{p}$ given in the left column.
}
\label{tab:pump_terms}
\begin{tabular}{C{2.5cm} C{2.5cm} C{2.5cm}}
\\
$\omega_\text{p}  \simeq $       & prefactor  & interaction \\ \hline \\
\multicolumn{3}{c}{ Stark shift}\\\\
					&$-2A_\text{H}    |\xi_\text{p}|^2$&									$\hat a^\dagger\hat a$\\
					&$-\chi    |\xi_\text{p}|^2$&									$\hat b^\dagger\hat b$\\
\\\multicolumn{3}{c}{Heating interactions}\\\\
$(\omega_\text{H}+\omega_\text{L})/2$					&$-A_\text{H}^{\frac{3}{4}}A_\text{L}^{\frac{1}{4}}   (\tilde\xi_\text{p}^*)^2$&		$\hat{a}\hat{b}$\\
$\omega_\text{H}+2\omega_\text{L}$						&$-\chi\tilde\xi_\text{p}^*/2$&										$\hat{a}\hat b^2 $\\
$2\omega_\text{H}+\omega_\text{L}$					&$-A_\text{H}^{\frac{3}{4}}A_\text{L}^{\frac{1}{4}}\tilde\xi_\text{p}^*$&			$\hat a^2\hat{b}$\\
\\\multicolumn{3}{c}{Cooling interactions}\\\\
$(\omega_\text{H}-\omega_\text{L})/2$					&$-A_\text{H}^{\frac{3}{4}}A_\text{L}^{\frac{1}{4}}   (\tilde\xi_\text{p}^*)^2$&		$\hat{a}\hat{b}^\dagger$\\
$\omega_\text{H}-2\omega_\text{L}$						&$-\chi\tilde\xi_\text{p}^*/2$&										$\hat{a}(\hat b^\dagger)^2$\\
$2\omega_\text{H}-\omega_\text{L}$					&$-A_\text{H}^{\frac{3}{4}}A_\text{L}^{\frac{1}{4}}\tilde\xi_\text{p}^*$&			$\hat a^2 \hat{b}^\dagger$\\
\\\multicolumn{3}{c}{Unused interactions}\\\\
$3\omega_\text{H}$					&$- A_\text{H}\tilde\xi_\text{p}/3$&										$\hat a^3$\\
$\omega_\text{H}/3$		 			&$-A_\text{H} (\tilde\xi_\text{p}^*)^3/3$&									$\hat a$\\
$\omega_\text{H}$					&$-A_\text{H}   (\tilde\xi_\text{p}^*)^2/2$&									$\hat a^2 $\\
$\omega_\text{H}$					&$-\chi\tilde\xi_\text{p}$&										$\hat{a}\hat b^\dagger\hat b $\\
$\omega_\text{H}$					&$-A_\text{H}\tilde\xi_\text{p}$&											$\hat a^\dagger\hat a^2$\\
$\omega_\text{H}$					&$- A_\text{H}(\tilde\xi_\text{p}^*)^3$&										$\hat{a}$\\
$\omega_\text{H}$					&$-A_\text{H}\tilde\xi_\text{p}$&											$\hat a$\\
$\omega_\text{H}$					&$-\chi\tilde\xi_\text{p}/2$&										$\hat{a}$\\
\\
$3\omega_\text{L}$					&$-A_\text{H}^{\frac{1}{4}}A_\text{L}^{\frac{3}{4}}\tilde\xi_\text{p}/3$&		$\hat b^3$\\
$\omega_\text{L}/3$					&$-A_\text{H}^{\frac{3}{4}}A_\text{L}^{\frac{1}{4}}(\tilde\xi_\text{p}^*)^3/3$&		$\hat{b}$\\
$\omega_\text{L}$					&$-\chi    \tilde\xi_\text{p}^2/4$&								$\hat b^2 $\\
$\omega_\text{L}$					&$-2A_\text{H}^{\frac{3}{4}}A_\text{L}^{\frac{1}{4}}\tilde\xi_\text{p}$&			$\hat a^\dagger\hat a\hat{b}$\\
$\omega_\text{L}$					&$-A_\text{H}^{\frac{1}{4}}A_\text{L}^{\frac{3}{4}}\tilde\xi_\text{p}$&			$\hat b^\dagger\hat b^2$\\
$\omega_\text{L}$					&$-A_\text{H}^{\frac{3}{4}}A_\text{L}^{\frac{1}{4}}(\tilde\xi_\text{p}^*)^3$&		$\hat{b}$\\
$\omega_\text{L}$					&$-A_\text{H}^{\frac{1}{4}}A_\text{L}^{\frac{3}{4}}\tilde\xi_\text{p}$&			$\hat b$\\
$\omega_\text{L}$					&$-A_\text{H}^{\frac{3}{4}}A_\text{L}^{\frac{1}{4}}\tilde\xi_\text{p}$&			$\hat{b}$\\

\end{tabular}
\end{table}

We now move to the interaction picture through the unitary transformation 
\begin{equation}
\begin{split}
	U_\text{int}&=e^{i \hat{H}_{4,\text{diag}}t/\hbar}\ ,
\end{split}
\end{equation}
$\hat{H}_{4,\text{diag}}$ is diagonal in the Fock state basis $\left\{\ket{j,n}\right\}_{\substack{n=0,1,2,.. \\ j=g,e,f,..}}$
\begin{equation}
\begin{split}
	\hat{H}_0&=\sum_{\substack{n=0,1,2,.. \\ j=g,e,f,..}}\hbar\epsilon_{j,n}\ket{j,n}\bra{j,n}\ ,\\
	\text{where }\epsilon_{j,n}&= n\omega_\text{L} - \frac{A_\text{L}}{2\hbar}\left(n^2-n\right)\\
	&+j\omega_\text{H} - \frac{A_\text{H}}{2\hbar}\left(j^2-j\right)\\
	&-nj\chi/\hbar\ .
\end{split}
\end{equation}
To determine $\hat H_\text{p}$ in this frame, it suffices to know the expression of annihilation operators in this frame.
We will take as an example the term we use for cooling, which reads in the interaction picture
\begin{equation}
\begin{split}
	&U_\text{int}\left(-A_\text{H}^{\frac{3}{4}}A_\text{L}^{\frac{1}{4}}   (\tilde\xi_\text{p}^*)^2\hat{a}^2\hat b^\dagger\right) U_\text{int}^\dagger +h.c. \\
	= -A_\text{H}^{\frac{3}{4}}&A_\text{L}^{\frac{1}{4}}   (\tilde\xi_\text{p}^*)^2(U\hat{a}U^\dagger)^2 (U\hat bU^\dagger)^\dagger +h.c.\ .
\end{split}
	\label{eq:cooling_out_of_interaction_picture}
\end{equation}
Since $\hat H_0$ is diagonal, exponentiating it only requires exponentiating each of the diagonal elements, and the annihilation operators in the interaction picture are
\begin{equation}
\begin{split}
	U_\text{int}\hat{a}U_\text{int}^\dagger &=\sum_{\substack{n=0,1,.. \\ j=g,e,..}} \sqrt{j+1}e^{-(\epsilon_{n,j+1} - \epsilon_{n,j})t/\hbar}\ket{j,n}\bra{j+1,n}\\
	U_\text{int}\hat{b}U_\text{int}^\dagger &=\sum_{\substack{i=0,1,.. \\ j=g,e,..}} \sqrt{n+1}e^{-(\epsilon_{j,n+1} - \epsilon_{j,n})t/\hbar}\ket{j,n}\bra{j,n+1}\ .
\end{split}
\label{eq:ab_interaction_picture}
\end{equation}
Note that if the system were harmonic, these expressions would simplify to $e^{-i\omega_\text{H} t}\hat a$ and $e^{-i\omega_\text{L} t}\hat b$.
If we substitute Eqs.~(\ref{eq:ab_interaction_picture}) into Eq.~(\ref{eq:cooling_out_of_interaction_picture}), one of the terms we obtain is
\begin{equation}
\begin{split}
	-\hbar ge^{i\left(\omega_\text{p}-\left(2\omega_\text{H}-A_\text{H}/\hbar-\omega_\text{L}\right)\right)t}&\ket{g,1}\bra{f,0} +h.c.\ ,
\end{split}
\end{equation}
where we defined the interaction strength
\begin{equation}
	g=\sqrt{2}A_\text{H}^{\frac{3}{4}}A_\text{L}^{\frac{1}{4}}|\xi_\text{p}|/\hbar\ .
\end{equation}
By choosing the pump frequency $\omega_\text{p} = 2\omega_\text{H} - A_\text{H}/\hbar - \omega_\text{L} $, the term becomes time-independent, making it more relevant than the other terms of $\hat H_P$ as we will derive next.
More generally, we can engineer the cooling interactions
\begin{equation}
\begin{split}
	&-\hbar g\sqrt{n+1}\ket{f,n}\bra{g,n+1} +h.c.\ ,
\end{split}
\label{eq:cooling_interaction_n}
\end{equation}
by choosing the pump frequencies
\begin{equation}
	\omega_\text{p} = 2\omega_\text{H}-2n\chi/\hbar - A_\text{H}/\hbar - \omega_\text{L}\ .
\end{equation}

This is the interaction used in all expriments presented in the last three figures of the main text.
Cooling by driving the $|g,1\rangle\leftrightarrow|e,0\rangle$ transition may seem like a more natural choice, but it is a two pump-photon process (due to four-wave mixing selection rules), and hence requires higher pumping power. 
Additionally, due to its higher energy, the $|f,0\rangle$ state has a lower thermal occupation than $|e,0\rangle$.
As discussed below, high pump powers and thermal occupation of the qubit place strong limitations on the cooling efficiency.

Rather than lowered, the number of excitations in the low mode can also be raised using interactions of the form 
\begin{equation}
\begin{split}
	&-\hbar g\sqrt{n+1}\ket{f,n+1}\bra{g,n} +h.c.\ ,
	\label{eq:raising_interaction}
\end{split}
\end{equation}
which are realized by choosing the pump frequencies
\begin{equation}
	\omega_\text{p} = 2\omega_\text{H}-2(n+1)\chi/\hbar - A_\text{H}/\hbar + \omega_\text{L}\ .
\end{equation}
\subsubsection{Derivation of cooling rate}
\label{sec:cooling_rate}

In this section we focus on the cooling interaction of Eq.~(\ref{eq:cooling_interaction_n}), however the methodology described is generalizable to all interaction terms.
The objective of this section is to translate the interaction term derived previously into a cooling rate for the low mode.
We assume that this interaction is sufficiently weak to enable us to perform first-order perturbation theory, considering the high mode as a fluctuating quantum noise source $\hat{F}_H$ perturbing the low mode following App.~B.1 of Ref.~\cite{clerk2010introduction}.
An initial state of the low mode $|n\rangle$ will evolve following
\begin{equation}
\begin{split}
    |\psi(t)\rangle=|n\rangle
    +i\sqrt{n}g\left(\int_0^t d\tau e^{i\Delta \tau}\hat{F}_H (\tau) \ket{n-1}\bra{n}\right)|n\rangle\
\end{split}
\end{equation}
where $\hat{F}_H (\tau) = \left(\ket{f}\bra{g}\right) (\tau)$ is treated as an independent noise source acting on the Hilbert space of the high mode. 
We consider the transition is off-resonantly driven such that the time-dependence in the interaction picture is not completely eliminated and the interaction term rotates at 
\begin{equation}
	\Delta = \omega_\text{p} - \left(2\omega_\text{H}-2n\chi/\hbar - A_\text{H}/\hbar - \omega_\text{L}\right)\ ,
\end{equation}
The probability amplitude of finding the low mode  in $|n-1\rangle$ is
\begin{equation}
    \langle n-1|\psi(t)\rangle =i\sqrt{n}g\int_0^t d\tau e^{i\Delta \tau}\left(\ket{f}\bra{g}\right) (\tau) \ ,
\end{equation}
leading to a probability
\begin{equation}
\begin{split}
    &|\langle n-1|\psi(t)\rangle|^2 =\langle n-1|\psi(t)\rangle^\dagger \langle n-1|\psi(t)\rangle\\
    &=ng^2\int_0^t\int_0^t d\tau_1 d\tau_2 e^{i\Delta (\tau_2-\tau_1)} \left(\ket{f}\bra{g}\right)^\dagger (\tau_1)\left(\ket{f}\bra{g}\right) (\tau_2)\ .
    \label{proba}
\end{split}
\end{equation}
%
%
Note that $|\langle n-1|\psi(t)\rangle|^2$ is still a quantum operator acting on the high-mode Hilbert space.
To obtain a classical probability, we now calculate its expectation value $\langle . \rangle_H$, provided that the high mode evolves in steady-state under thermal effects and dissipation
\begin{equation}
\begin{split}
    &p_{n\rightarrow n-1}(t)=\langle|\langle n-1|\psi(t)\rangle|^2\rangle_H\\
    &=ng^2\int_0^t\int_0^t d\tau_1 d\tau_2 e^{i\Delta (\tau_2-\tau_1)}\langle\left(\ket{f}\bra{g}\right)^\dagger (\tau_1)\left(\ket{f}\bra{g}\right) (\tau_2)\rangle_H\ .
\end{split}
\end{equation}
As in Appendix A.2 of \cite{clerk2010introduction}, we transform the double integral $S$ to
\begin{equation}
\begin{split}
    S &= \int_0^t d\tau_1 \int_0^t d\tau_2 e^{i\Delta (\tau_2-\tau_1)}\langle\left(\ket{g}\bra{f}\right) (\tau_1)\left(\ket{f}\bra{g}\right) (\tau_2)\rangle_H\\
    &=\int_0^t dT \int_{-B(T)}^{B(T)} d\tau e^{-i\Delta \tau}\langle\left(\ket{g}\bra{f}\right) (T + \tau/2)\\
    &\ \ \ \ \ \ \ \ \ \ \ \ \ \ \ \ \ \ \ \ \ \ \ \ \ \ \ \ \ \ \ \ \times\left(\ket{f}\bra{g}\right) (T - \tau/2)\rangle_H\ ,\\
	&\text{where }B(T) = 2T\text{ if }T< t/2\\
	        &\ \ \ \ \ \ \ \ \ \ \ \ \ \ \ \ = 2(t-T)\text{ if }T> t/2
\end{split}
\end{equation}
For time-scales larger than the decay rate of the high mode $\tau\gg 1/\kappa$, the two time-dependent high-mode operators are not correlated and the integrand will vanish (see Appendix A.2 of \cite{clerk2010introduction}). 
We can therefore extend the range of the inner integral to $\pm \infty$ in estimating the probability at a time $t\gg1/\kappa$. 
\begin{equation}
\begin{split}
	S =\int_0^t dT \int_{-\infty}^{+\infty} d\tau e^{-i\Delta \tau}\langle&\left(\ket{g}\bra{f}\right) (T + \tau/2)\\
    \times&\left(\ket{f}\bra{g}\right) (T - \tau/2)\rangle_H\ .\\
\end{split}
\end{equation}
Using time-translation invariance, we can remove the dependence on $T$
\begin{equation}
\begin{split}
	S &=\int_0^t dT \int_{-\infty}^{+\infty} d\tau e^{-i\Delta \tau} \langle\left(\ket{g}\bra{f}\right) (\tau)\left(\ket{f}\bra{g}\right) (0)\rangle_H\\
	&=t\int_{-\infty}^{+\infty} d\tau e^{-i\Delta \tau}\langle\left(\ket{g}\bra{f}\right) (\tau)\left(\ket{f}\bra{g}\right) (0)\rangle_H\ ,
\end{split}
\end{equation}
such that the rate becomes time-independent
\begin{equation}
\begin{split}
	\Gamma_{n\rightarrow n-1} &= p_{n\rightarrow n-1}(t)/t\\
		&=ng^2\int_{-\infty}^{+\infty} d\tau e^{-i\Delta \tau} \langle\left(\ket{g}\bra{f}\right) (\tau)\left(\ket{f}\bra{g}\right) (0)\rangle_H\ .\\
\end{split}
\end{equation}
Using time-translation invariance, we find that for negative values of $\tau$,
\begin{equation}
\begin{split}
\langle\left(\ket{g}\bra{f}\right)& (-|\tau|)\left(\ket{f}\bra{g}\right) (0)\rangle_H\\
 & = \langle\left(\ket{g}\bra{f}\right) (0)\left(\ket{f}\bra{g}\right) (|\tau|)\rangle_H \\
 & = \langle\left(\ket{g}\bra{f}\right) (|\tau|)\left(\ket{f}\bra{g}\right) (0)\rangle^*_H\ ,
\end{split}
\end{equation}
leading to 
\begin{equation}
\begin{split}
	\Gamma_{n\rightarrow n-1} &=ng^2\int_{0}^{\infty} d\tau e^{-i\Delta \tau}\langle\left(\ket{g}\bra{f}\right) (\tau)\left(\ket{f}\bra{g}\right) (0)\rangle_H\\
	&+ng^2\int_{0}^{\infty} d\tau e^{-i\Delta \tau}\langle\left(\ket{g}\bra{f}\right) (\tau)\left(\ket{f}\bra{g}\right) (0)\rangle^*_H\\
	=2ng^2 &\text{Re}\left(\int_{0}^{\infty} d\tau e^{-i\Delta \tau} \langle\left(\ket{g}\bra{f}\right) (\tau)\left(\ket{f}\bra{g}\right) (0)\rangle_H\right)\ .\\
\end{split}
\end{equation}
In the steady state of the system, the quantum regression theorem 
can be shown to reduce the expression to
\begin{equation}
\Gamma_{n\rightarrow n-1} = 2ng^2\text{Re}\left(\int_{0}^{\infty} d\tau  e^{-i\Delta \tau}~\text{Tr}\left[\ket{g}\bra{f}e^{\mathcal{L}\tau}\ket{f}\bra{g}\hat\rho\right]\right)\ ,
\end{equation}
%
where $\hat\rho$ is the steady-state density matrix of the high mode and $e^{\mathcal{L}t}$ its propagator, a function which takes a density matrix as an input and evolves it up to a time $t$ following the Lindblad equation.
Reducing the high mode to a three-level system and considering dissipation and thermal effects, this trace can be calculated analytically using the QuantumUtils Mathematica library
\begin{equation}
	\text{Tr}\left[\ket{g}\bra{f}e^{\mathcal{L}t}\ket{f}\bra{g}\hat\rho\right]=P_ge^{-\kappa t(1+\frac{3}{2}n_\text{th}^{(H)})}\ .
\end{equation}
By only considering dissipation and thermalization, we made the assumption that an excitation could not be driven back from $\ket{f,n-1}$ to $\ket{g,n}$ under the effect of pumping, \textit{i.e. }we assume $2\kappa\gg \sqrt{n}g$, that we are far from the strong coupling regime.
After integration, we obtain
\begin{equation}
	\Gamma_{n\rightarrow n-1} = \frac{2ng^2P_g}{\kappa(1+\frac{3}{2}n_\text{th}^{(H)})}\frac{1}{1+\left(\frac{\Delta}{\kappa}\right)^2}\ .
	\label{eq:rate_down}
\end{equation}
Following the same method, we also obtain for the hermitian conjugate of this interaction term
\begin{equation}
	\Gamma_{n-1\rightarrow n} = \frac{2ng^2P_f}{\kappa(1+\frac{3}{2}n_\text{th}^{(H)})}\frac{1}{1+\left(\frac{\Delta}{\kappa}\right)^2}\ ,
	\label{eq:rate_up}
\end{equation}
if the $\ket{f}$ level is populated, we find that there is a probability for the pump to raise the number of excitations in the low mode rather than lower it.
We refer to the steady state population of the ground and second-excited state of the high mode as $P_g$ and $P_f$ respectively.
The same calculation can be performed for the raising interaction, which yields identical rates only with $P_g$ and $P_f$ interchanged.
A good figure of merit of the cooling efficiency is then to compare this rate with $\gamma$, yielding the cooperativity
\begin{equation}
	C = \frac{\Gamma_{1\rightarrow 0}}{\gamma} =\frac{2g^2}{\kappa\gamma(1+\frac{3}{2}n_\text{th}^{(H)})}\frac{1}{1+\left(\frac{\Delta}{\kappa}\right)^2}\ .
	\label{eq:full_cooperativity}
\end{equation}

\subsubsection{Semi-classical description of the cooling process}
With the cooling rate above, we can construct a semi-classical set of rate equations describing the competition between thermalization and cooling.
They would correspond to the diagonal part of a Lindblad equation, and equates the population leaving and arriving to a given state of the low mode.
We restrict ourselves to the driving of $\ket{f,0}\leftrightarrow\ket{g,1}$ as in the experiment of Fig.~2, where these equations can be written as
\begin{equation}
	\dot P_0 = P_1\left(\gamma CP_g + \gamma (n_\text{th}+1)\right) - P_0\left(\gamma CP_f + \gamma n_\text{th}\right),
\end{equation}
\begin{equation}
\begin{split}
	\dot P_1 &= -P_1\left(\gamma CP_g + \gamma (n_\text{th}+1)\right) + P_0\left(\gamma CP_f + \gamma n_\text{th}\right)\\
	&  -2P_1\gamma n_\text{th} + 2P_2\gamma (n_\text{th}+1)\ ,
\end{split}
\end{equation}
and, for $n\ge2$
\begin{equation}
\begin{split}
	\dot P_n &= -n \gamma (n_\text{th}+1) P_n + n\gamma n_\text{th}P_{n-1}\\
	& -(n+1)P_n\gamma n_\text{th} + (n+1)P_{n+1}\gamma (n_\text{th}+1)\ .
\end{split}
\end{equation}
In steady state ($\dot P=0$), the solution is a function of $P_0$
\begin{equation}
\begin{split}
	\frac{P_0}{P_1}&=\frac{CP_g + n_\text{th}+1}{CP_f + n_\text{th}} =A\ ,\\
	\frac{P_n}{P_{n+1}}&=\frac{n_\text{th}+1}{n_\text{th}} = B\ \text{for}\ n\ge1\ .
\end{split}
	\label{eq:cooling_formula}
\end{equation}
We reach a unique solution by imposing $\sum_nP_n=1$, which yields an expression for $P_0$ 
\begin{equation}
	P_0 = \frac{
A
\left(A-1\right)
\left(B-1\right)
}{
B(A^2-1)
+ A(1-A)
}\ .
\label{eq:P0_with_thermal}
\end{equation}
This expression is used in Fig.~\ref{fig:S_version_fig_2} to show the temperature limited evolution of $P_0$ as a function of cooperativity.
A more accurate description of the cooling process at high cooperativities comes from a numerical simulation taking the strong coupling limit and off-resonant driving of other four-wave mixing processes into account.
%

\subsection{Limiting factors to cooling}

\begin{figure}[ht!]
\includegraphics[width=0.5\textwidth]{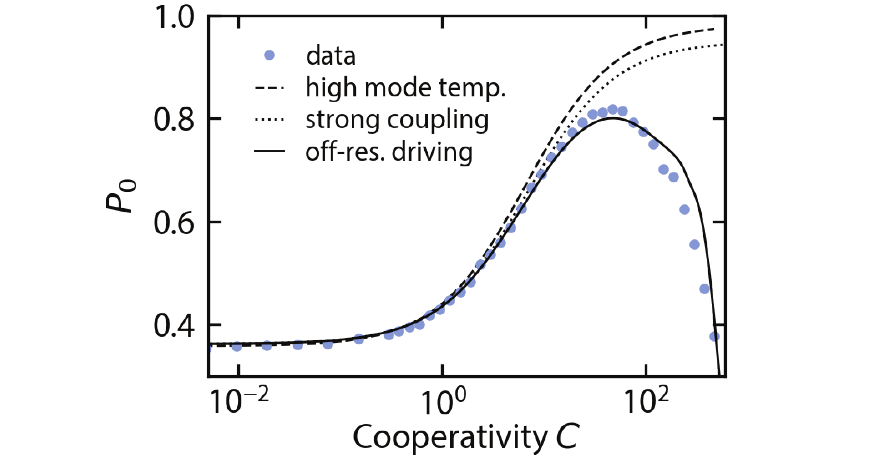}
\caption[Limiting factors to cooling.]{\textbf{Limiting factors to cooling. }
$P_0$ as a function of $C$, with dots showing data points identically to Fig.~2.
The decrease of $P_0$ at large $C$ is not captured by the cooling limitation due to thermal population of the $\ket{f}$ state (dashed line) or the limit imposed by strong coupling (dotted line), where the pump hybridizes the $\ket{g,1}$ and $\ket{f,0}$ states. 
The solid curve shows a prediction considering the off-resonant driving of other sideband transitions by the pump: as the cooling process starts to saturate due to the strong coupling limit, the driving rate of transitions that increase the photon number overpowers the cooling effect.
}
\label{fig:S_version_fig_2}
\end{figure}

Here, we discuss three limiting factors to the cooling experiment (Fig.~2), ending with some notes on how to improve the device cooling performance
The first limiting factor is the thermal occupation of the high mode. 
The pump tone drives the population from $|g,1\rangle$ to $|f,0\rangle$, but the reverse process also occurs since the $f$ level has a small thermal population $P_f\simeq0.006$ (see Sec.~\ref{sec:temperature}).
This leads to the limit $P_1/P_0 > P_f/P_g$ (dashed line in Fig.~\ref{fig:S_version_fig_2}) for which we have derived an exact analytical expression (Eq.~(\ref{eq:P0_with_thermal}))

The second limiting factor is that of strong coupling (similar to in optomechanical cooling~\cite{teufel2011circuit}), where the pump hybridizes the $|g,1\rangle$ and $|f,0\rangle$ states.
If $g$ exceeds the decay rate $2\kappa$, the population of state $|g,1\rangle$ will be driven to $|f,0\rangle$ and then transfered back to $|g,1\rangle$ without having the time to decay to $|e,0\rangle$.
To simulate this effect, we compute the steady state of the system by solving a Lindblad equation numerically (see ~\ref{sec:cooling_simulation}).
The result is shown as a dotted line in Fig.~\ref{fig:S_version_fig_2}, which additionally takes into account the population of the high mode.
As with the thermal effect, the strong coupling limit only imposes an upper bound on $P_0$, rather than predicting its decrease at high $C$.
%


%
When the cooling tone is detuned by $\Delta$ from its transition frequency, the cooperativity acquires a factor $1/\left(1+\Delta^2/\kappa^2\right)$ (Eq.~(\ref{eq:rate_up})).
A similar formula applies to all other four-wave mixing processes, including raising interactions (Eq.~(\ref{eq:raising_interaction})).
If the latter are far-detuned, their off-resonant driving will have little impact on the system.
However, as the cooling process starts to saturate due to the previously discussed limiting factors, the driving of other transitions is still far from saturation and can overpower the cooling effect.
What ensues is a competition between off-resonantly driven transitions that cool and raise the photon occupation.
We simulate this by following the bootstrap step of the adaptive rotating-wave approximation method of Ref.~\cite{baker2018adaptive}, which offers a way to include the most relevant off-resonantly driven transitions to the system Hamiltonian (see Sec.~\ref{sec:cooling_simulation}).
The result is shown as the solid curve of Fig.~\ref{fig:S_version_fig_2} which predicts the maximum $P_0$ and the strong cooperativity behavior.
We emphasize that, except for a small shift on the calibrated cooperativity-axis, the theoretical curves do not correspond to a fit to the data, but rather constitute a prediction based on the independently determined dissipation rates, thermal occupations and circuit parameters.
From this simulation we extract that, at maximum $P_0=0.82$, the average photon number in the cooled resonator is $\bar n =0.65$.
Note that 
\begin{equation}
\begin{split}
	\bar n &= 0\times(P_{g0}+P_{e0}+P_{f0}+...)\\
	&+1\times(P_{g1}+P_{e1}+P_{f1}+...)\\
	&+2\times(P_{g2}+P_{e2}+P_{f2}+...)\\
	&+...
\end{split}
\end{equation}
The first 10 most populated levels are: $P_{g0} = 0.736$, $P_{e0} = 0.067$, $P_{g1} = 0.036$, $P_{g2} = 0.028$, $P_{g3} = 0.028$, $P_{g4} = 0.022$, $P_{g5} = 0.017$, $P_{g6} = 0.014$, $P_{g7} = 0.011$, $P_{f0} = 0.009$, where  $P_{j,n}$ refers to the occupation of state $\ket{j,n}$.
Taking only the contribution of these states into account in the above formula already leads to $\bar n = 0.51$, and including the occupation of all 50 simulated levels leads to $\bar n =0.65$.


Determining the ideal system parameters to improve cooling (and Fock-state stabilization fidelity) is not straightforward.
One path to improvement could lie in determining values of $A_H$ and $\chi$ which minimize the effect of off-resonant driving by moving the most problematic transitions away from the cooling frequency.
Another is to reach a higher ground-state occupation before being limited by strong coupling, which can only be achieved by reducing the resonators dissipation $\gamma$.
Decreasing the high mode dissipation $\kappa$ is not necessarily beneficial: it diminishes off-resonant driving, but strong coupling would occur at smaller pump powers.
For our system, decreasing $\kappa$ in the simulation of Fig.~2C results in a lower ground-state occupation.

\clearpage
\onecolumngrid
\section{Numerical procedures}
\twocolumngrid
\subsection{Spectrum}

The eigenfrequencies of the system are determined by diagonalizing the system Hamiltonian.
%
%
Unless specified otherwise, we diagonalize the Hamiltonian of Eq.~\ref{eq:Hamiltonian_8th_order} with the junction non-linearity Taylor expanded to 8-th order.
We consider 10 excitations in the high mode and 20 in the low mode, and have verified that extending the Hilbert space further only leads to negligeable changes in the obtained spectrum.
This diagonalization also provides the dressed eigenstates $\ket{j,n}$, which are to be distinguished from the bare eigenstates $\widetilde{\ket{j,n}}_{\substack{n=0,1,2,.. \\ j=g,e,f,..}}$.

\subsection{Microwave reflection}
\label{sec:numerics_S11}

In order to compute the microwave reflection of the device, we solve a Lindblad equation using Qutip~\cite{johansson2012qutip}.
The Hamiltonian is written in the dressed basis defined above, it is hence diagonal with entries corresponding to the eigenfrequencies obtained in the diagonalization.
We consider 5 high-mode excitations and 10 low-mode excitations.
We add the drive term $i\hbar\epsilon_d(\hat a^\dagger -\hat a)$ defined in the dressed basis, and move to the frame rotating at the drive frequency $\omega_d$ by adding $-\hbar\omega_d\hat a ^\dagger \hat a$.
We add jump operators defined in the dressed basis by
\begin{equation}
\begin{split}
	(n_\text{th}^{(H)}\kappa)^{\frac{1}{2}}\hat a^\dagger\ ,\ ((n_\text{th}^{(H)}+1)\kappa)^{\frac{1}{2}}\hat a\ ,\\
	(n_\text{th}\gamma)^{\frac{1}{2}}\hat b^\dagger\ , \ ((n_\text{th}+1)\gamma)^{\frac{1}{2}}\hat b\ ,
	\label{eq:collapse_ops}
\end{split}
\end{equation}
to describe dissipation and thermal effects.
Finally, we compute the expectation value of $\hat S_{11}=1-\frac{\kappa_\text{ext}}{\epsilon_\text{d}}\hat{a}$ for different drive frequencies.
As shown in Fig.~\ref{fig:S_temperature}, this computation is in excellent agreement with the sum of Lorentzian formula of Eq.~(\ref{eq:sum_of_lorentzians}).

\subsection{Cooling simulation}
\label{sec:cooling_simulation}

We use a similar method for the adaptive rotating-wave approximation (aRWA) simulation of Fig.~2.
We start with the same diagonal Hamiltonian.
We denote by $\omega_{j,n}$ the eigenfrequency of the dressed eigenstates $\ket{j,n}$.
As a result of the collapse operators of Eq.~(\ref{eq:collapse_ops}), a dressed state of the system $\ket{j,n}$ will have a total decay rate to other states of the system 
\begin{equation}
\begin{split}
	\Gamma_{j,n} = (j+1)(n_\text{th}^{(H)}\kappa) +j((n_\text{th}^{(H)}+1)\kappa)\\ +(n+1)(n_\text{th}\gamma) +n((n_\text{th}+1)\gamma)\ .
\end{split}
\end{equation}
Following Ref.~\cite{baker2018adaptive}, we can then estimate the impact of a pump tone at a frequency $\omega_p$ and driving rate $\epsilon_p$ on the steady state of the system.
Two states $\ket{k}=\ket{j,n}$ and $\ket{k'}=\ket{j',n'}$ will be coupled by this pump.
And to first order in $\epsilon_p$, the only change in the steady state density matrix will be in its off-diagonal element 
\begin{equation}
	\rho_{kk'} = \frac{V_{kk'}(P_{k'}-P_k)}{(\omega_{k'}-\omega_k)-\omega_p+i(\Gamma_k+\Gamma_{k'})/2}\ ,
	\label{eq:ranking_parameter}
\end{equation}
where $P_{k}$ is the occupation of state $\ket{k}$ under the collapse operators of Eq.~(\ref{eq:collapse_ops}).
The dipole moment $V_{kk'} = \bra{k}\epsilon_p(\tilde a + \tilde a^\dagger)\ket{k'}$ is computed using annihilation $\tilde a$ and creation operators $\tilde a^\dagger$ defined in the bare basis. 
The transitions between all the states are then ranked with decreasing $|\rho_{kk'}|$ (\textit{i.e.} decreasing relevance).
The most relevant terms are added in the form $\hbar V_{kk'}\ket{k}\bra{k'}$ to the Hamiltonian which is moved to the rotating frame in which states $\ket{k}$ and $\ket{k'}$ are resonant.

In Fig.~\ref{fig:S_version_fig_2}, we perform this calculation for $\omega_p = \omega_{f,0}-\omega_{g,1}$.
We show both the result of including a maximum number of transitions (465) and a single transition.
It was only possible to include 465 transitions out of the 650 transitions which have a non zero dipole moment.
This is due to limitations in the construction of the rotating frame, for more details see Ref.~\cite{baker2018adaptive}.

\begin{figure*}[ht!]
\includegraphics[width=0.75\textwidth]{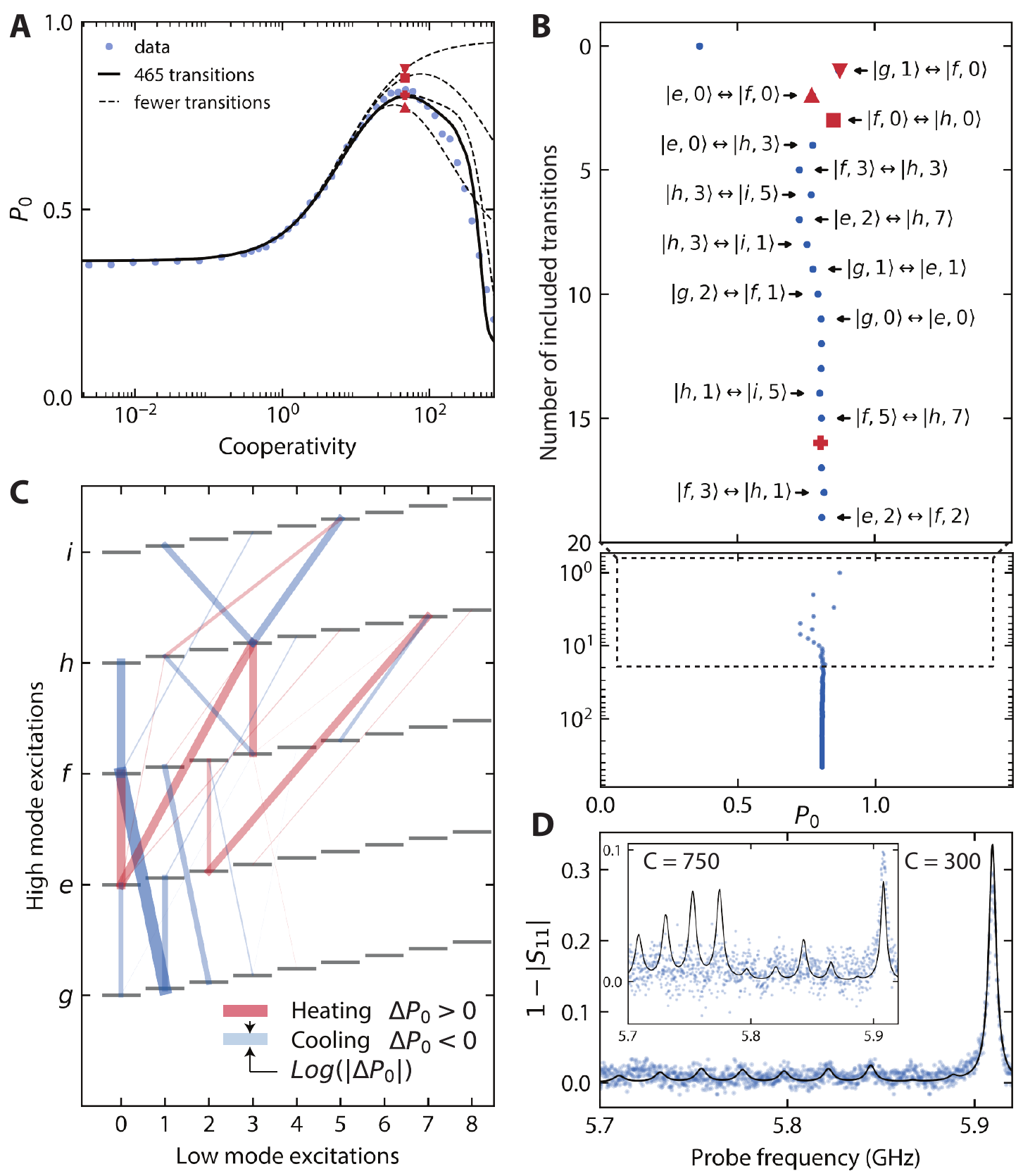}
\caption[Simulation of off-resonantly driving transitions in cooling experiment.]{\textbf{Simulation of off-resonantly driving transitions in cooling experiment. }
\textbf{A: }Ground-state occupation $P_0$ as a function of cooperativity in the cooling experiment of Fig.~2.
We show data (blue dots) and a simulation including the driving of 1, 2, 3, 16 (dashed curves) and 465 transitions (solid curve) using the adaptive rotating-wave approximation~\cite{baker2018adaptive} (aRWA).
The red symbols in panels A and B correspond to identical simulation points.
\textbf{B: }
Evolution of $P_0$ at a fixed cooperativity $C=46$.
For each point, we consider an additional transition.
\textbf{C: }
The transitions leading to the largest change in $P_0$ are displayed in the system energy diagram.
The transitions are colored red (blue) if adding them causes an increase (decrease) of $P_0$.
This distinction should be interpreted with care since the change in $P_0$ may be the result of multiple transitions interacting.
The thickness of the lines is logarithmically related to the change in $P_0$ that comes from adding the transition.
\textbf{D: }
Experimental spectrum (blue dots) and numerical predictions (sum of Lorentzian formula Eq.~(\ref{eq:sum_of_lorentzians})) at very high cooling powers.
We use the aRWA simulation to estimate the amplitudes of each Lorentzian peak. 
Up to a cooperativity of $C=300$, the data is consistent with aRWA predictions.
However, there is a clear deviation between data and simulation at the highest powers (inset).
}
\label{fig:S_tree}
\end{figure*}

In Fig.~\ref{fig:S_tree}, we study how each transition affects the steady-state of the system.
Ranking using  $|\rho_{kk'}|$ does not take into account that multiple transitions may interact.
To rank the relevance of the transitions in a more realistic way, we further rank the transitions following their impact on $P_0$.
We add transitions one by one in the simulation, recording for each transition the change $\Delta P_0$ that ensues.
We then rank the transitions with decreasing $|\Delta P_0|$, and repeat: we add the transitions in the new order one by one, rank them and start again until reaching convergence.

This simulation is in good agreement with the data except at the very highest powers (see Fig.~\ref{fig:S_tree}D).
There are four possible limitations in our aRWA simulation that could be the cause of this discrepancy.
First, our implementation of aRWA does not take into account the AC-Stark shift of each level.
Present only at high powers, these AC-Stark shifts could bring certain transitions in or out of resonance with each-other, modifying the final steady-state of the system.
Secondly, we work with a Hilbert space of only 10 excitations in the low mode.
At the highest power, the simulation indicates an average low-mode occupation of $~5$ and a larger Hilbert space may be needed to reach more accurate results.
Thirdly, only first order transitions were considered in the ranking of the transitions, so no higher order processes, such as those shown in Fig.~\ref{fig:S_all_transitions}C, are taken into account.
Fourthly, we rank transitions with Eq.~(\ref{eq:ranking_parameter}) using $P_{k}$ the occupation of states $\ket{k}$ under the collapse operators of Eq.~(\ref{eq:collapse_ops}).
However $P_{k}$ may change under the effect of the driving, modifying the relevance of a given transition.
This can be taken into account as described in Ref.~\cite{baker2018adaptive}, but is too computationally expensive with the Hilbert-space size used here.

\clearpage
\onecolumngrid
\section{Background subtraction}
\twocolumngrid
\subsection{Network analysis}

Most of our data analysis relies on fitting a sum of complex Lorentzians (see Eq.(~\ref{eq:sum_of_lorentzians})), to the measured microwave reflection $S_{11}$ in both phase and amplitude.
The signal we acquire is affected by the imperfections of the microwave equipment used to carry the signals to and from the device.

These can be modeled by a two port network with $s$ parameters $s_{11},s_{22}$, corresponding to the reflections at the VNA ports (reflected back to the VNA) and at the device (reflected back to the device) respectively, and $s_{21},s_{12}$, corresponding to the attenuation chain from the VNA to the device and the amplification chain from the device to the VNA respectively.
\begin{figure}[ht!]
\includegraphics{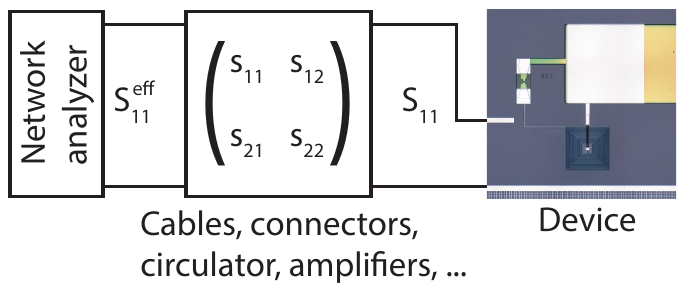}
\caption[Effective microwave network]{\textbf{Effective microwave network. }
We do not directly have access to the reflection at our device $S_{11}$.
We measure an effective reflection coefficient $S_{11}^{\text{eff}}$, affected by the imperfect microwave equipment between the network analyzer and device described by an $s-$matrix.
}
\label{fig:background_substraction}
\end{figure}
%

We hence measure with our VNA the effective microwave reflection
\begin{equation}
	S_{11}^{\text{eff}} = s_{11}+\frac{s_{12}s_{21}}{1-s_{22}S_{11}}S_{11}
\end{equation}
Note that these $s$ parameters are generally frequency dependent.
We make the approximation $s_{11},s_{22}\ll s_{12},s_{21}$, meaning we attribute most of the measured microwave background to the frequency dependent transmission of the attenuation and amplification chain.
The signal we want to measure is now proportional to a so-called ``microwave background''
\begin{equation}
	S_{11}^{\text{eff}} \simeq s_{12}s_{21}S_{11}\ ,
\end{equation}
which we have to experimentally measure.

\begin{figure}[ht!]
\includegraphics[width=0.45\textwidth]{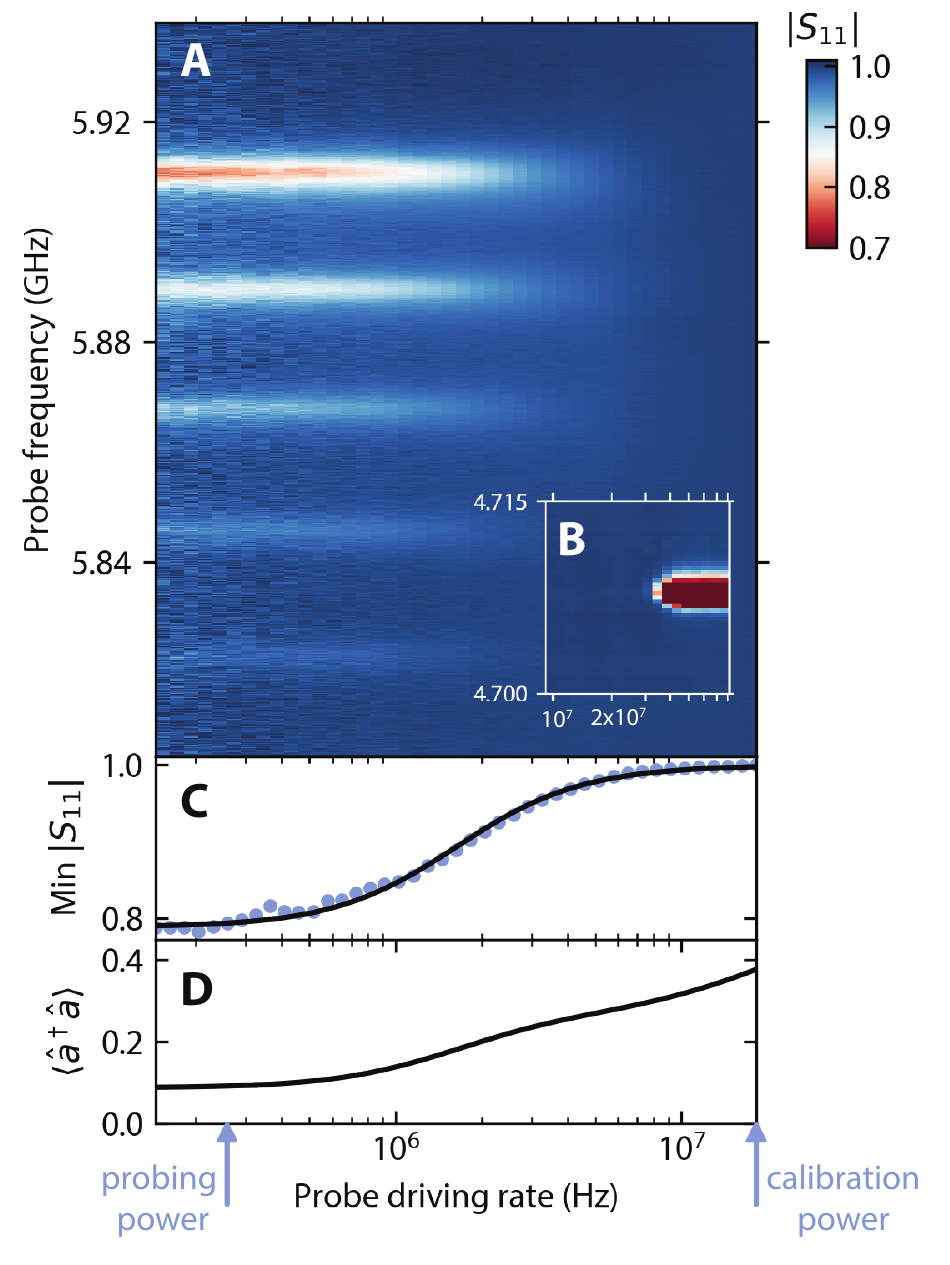}
\caption[Probe power dependence]{
	\textbf{High-probe-power behavior. A: }
	$|S_{11}|$ as a function of probe frequency and probe power. 
	\textbf{B: }
	At higher powers, the system starts to resonate at a different frequency, corresponding to the junction being replaced by an open circuit.
	\textbf{C: }
	Depth of the $n=0$ peak extracted from data (blue dots) and numerical steady-state calculation (see Sec.~\ref{sec:numerics_S11}).
	As the probe driving rate exceeds $\kappa$, the peaks vanish.
	We use the disappearance of peaks at the high power indicated by an arrow to acquire a microwave background that is subtracted (divided) in phase (amplitude) from all datasets.
	\textbf{D: }
	Population in the high mode as a function of probe power as extracted from simulation.
	We used this information to choose the probing power indicated by an arrow for all other experiments. 
	It is as high as possible to increase signal to noise ratio, but low enough to not populate the high mode.
}
\label{fig:S_power}
\end{figure}

\subsection{Measuring the microwave background}

As shown in Fig.~\ref{fig:S_power}, when probing the system at high power the device response is $S_{11}=1$, allowing us to extract the microwave background $s_{12}s_{21}$.
This phenomenon is a consequence of super-splitting as explained in \cite{Bishop2009a}, which we will briefly summarize here.

To understand super-splitting, we have to truncate the high mode to a two-level system constituted of its two first levels $\ket{g}$ and $\ket{e}$.
In the Bloch sphere, the probe tone will cause rotations around the y-axis and $1-S_{11}$ corresponds to the projection of the state vector on the x-axis.
For driving rates faster than $\kappa$, the state vector will rapidly rotate around the y-axis yielding a zero projection on the x-axis hence $S_{11}=1$ and no peak.
For driving rates slower than $\kappa$, random decays of the state vector will be very likely to occur before the state vector can rotate around the y-axis, yielding a non zero projection on the x-axis and a dip in the microwave reflection.
A signature of this effect is the splitting of the absorption peak in two for large probe powers.
Whilst our signal to noise does not allow the resolution of this feature, it is present in the fitted simulation, supporting this explanation.

At even higher power, the system starts to resonate at a different frequency, corresponding to the junction being replaced by an open circuit when the current traversing the junction exceeds the critical current.
%
This effect is shown in the inset, Fig.~\ref{fig:S_power}B.

We use the disappearance of peaks at a high power indicated by the arrow ``calibrating power" to acquire a microwave background that is subtracted (divided) in phase (amplitude) to all datasets.
\begin{equation}
	\frac{S_{11}^{\text{eff}}}{s_{12}s_{21}} \simeq S_{11}\ .
\end{equation}
%


\onecolumngrid
\section{Fitting}
\twocolumngrid

Here, we summarize our fitting routine.
We start by extracting $\gamma$ from the time-domain data, which will be used in the formula for the linewidth $\kappa_n$ in all subsequent fits.
By fitting the microwave reflection $S_{11}$ to a sum of Lorentzians (see Eq.~(\ref{eq:sum_of_lorentzians})), we get access to the peaks linewidths and amplitudes which allows us to determine $\kappa$, $\kappa_\text{ext}$ and $n_\text{th}^{(H)}$.
By fitting $S_{11}$ to the eigenfrequencies obtained from a diagonalization of the Hamiltonian of Eq.~(\ref{eq:Hamiltonian_8th_order}), we determine the values of the circuit elements.
The occupation of the low mode $n_\text{th}$ is determined separately for each individual experiment.
Each step is detailed in the subsections below.

\subsection{Low-frequency mode dissipation}
\begin{figure*}[ht!]
\includegraphics[width=1\textwidth]{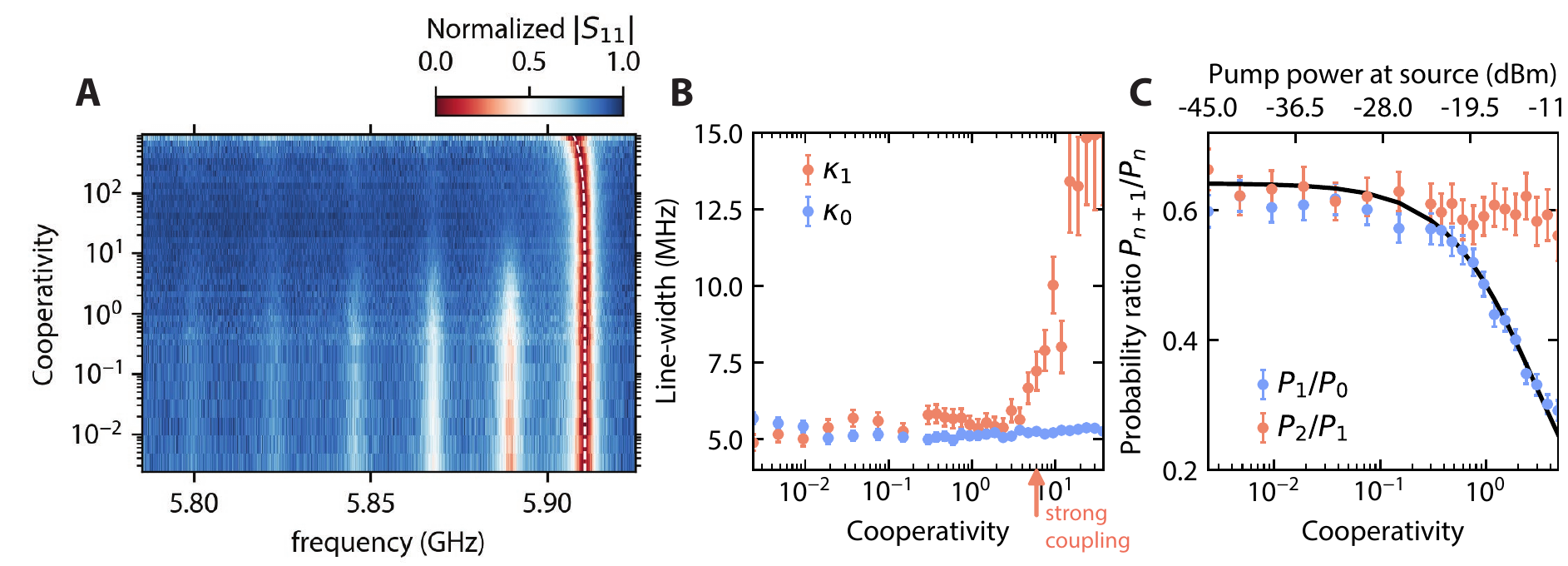}
\caption[AC-stark shift and cooperativity measurement]{\textbf{AC-stark shift and cooperativity measurement. A: }
	Normalized $|S_{11}|$ as a function of probe frequency and cooperativity in a single-pump cooling experiment.
	The AC-Stark shift of the $n=0$ peak follows the fitted white dashed line.
	%
	%
	\textbf{B: }
	Extracted line-widths of peaks $n=0,1$.
	Error bars correspond to the standard errors estimated from the least-squares fit.
	%
	%
	Fluctuations in $\kappa_n$ result from fluctuations in $n_\text{th}$.
	The onset of the strong-coupling regime, indicated by a red arrow, is seen through the increase in line-width of the $n=1$ peak.
	Below this value, we can accurately extract $P_1$ from the amplitude of the $n=1$.
	\textbf{C: }
	Ratio of extracted probabilities $P_2/P_1$ and $P_1/P_0$.
	The former is constant $P_2/P_1 = n_\text{th}/(1+n_\text{th})$, whilst the latter is fitted to $P_1/P_0 = (n_\text{th}+CP_f)/(1+n_\text{th}+CP_g)$, allowing us to convert pump power (top x-axis) to cooperativity (bottom x-axis).
}
\label{fig:S_cooling}
\end{figure*}

We start by fitting the thermalization from the ground-state measured in time-domain (Fig.~\ref{fig:S_time_domain}A) to determine $\gamma$.
Since the line-width of the $S_{11}(t)$ peaks is a function of $\gamma$ and $n_\text{th}$, we start by postulating these two values to extract a first estimate of the time evolution of $P_n$.
By fitting the evolution of $P_n$ to the rate equation of Eq.~(6), we extract a new value for $\gamma$ and $n_\text{th}$.
We then repeat this process many times, each time using the new values $\gamma$ and $n_\text{th}$ to fit $S_{11}(t)$, until we converge to $\gamma/2\pi=23.5\cdot 10^3s^{-1}$.
%
%

The low-frequency mode dissipation can also be measured without recourse to time-domain experiments.
The knowledge of the power dependent AC-stark shift and the cooperativity, measured in a single tone cooling experiment, is sufficient to extract $\gamma$.
We use this method to confirm our time-domain results, as well as verify the theory developed in Sec. ~\ref{sec:cooling_rate}.
First we measure the AC-stark shift of the $n=0$ peak, from which we extract the the proportionality factor $\xi^2/P$, between pumping rate $\xi$ and pump power $P$ (Fig.~\ref{fig:S_cooling}A).
Secondly we determine the power at which the strong coupling regime arises (Fig.~\ref{fig:S_cooling}B).
Above this power, the line-width of the $n=1$ peak will rise as the state $\ket{g,1}$ hybridizes with $\ket{f,0}$ under the effect of the cooling pump.
Below this power, the line-width of the $n=1$ peak is approximatively constant, and its height provides an accurate measure of $P_1$.
In this regime, we thirdly extract the ratio of probabilities $P_2/P_1$ and $P_1/P_0$
Following Eqs.~(\ref{eq:cooling_formula}), the former should remain constant $P_2/P_1 = n_\text{th}/(1+n_\text{th})$.
The latter, however, decreases with power, $P_1/P_0 = (n_\text{th}+CP_f)/(1+n_\text{th}+CP_g)$, and fitting this curve provides the conversion factor between cooperativity $C$ and power.
If we also know the anharmonicity $A_\text{H}$, cross-Kerr $\chi$, the high-mode occupation $n_\text{th}^{(H)}$ and dissipation rate $\kappa$, we can estimate the low-mode dissipation $\gamma = 2(\xi^2/P)/(C/P)A_\text{H}\chi/\kappa/(1+3n_\text{th}^{(H)}/2) \simeq 2\pi\cdot16\cdot 10^3s^{-1}$ close to the value obtained in time-domain.
The discrepancy is due to the inaccuracy of the relation $P_1/P_0 = (n_\text{th}+CP_f)/(1+n_\text{th}+CP_g)$, arising from the off-resonant driving of other four-wave mixing transitions.

\subsection{High-frequency mode dissipation and device temperature}
\label{sec:temperature}
\begin{figure*}[ht!]
\includegraphics{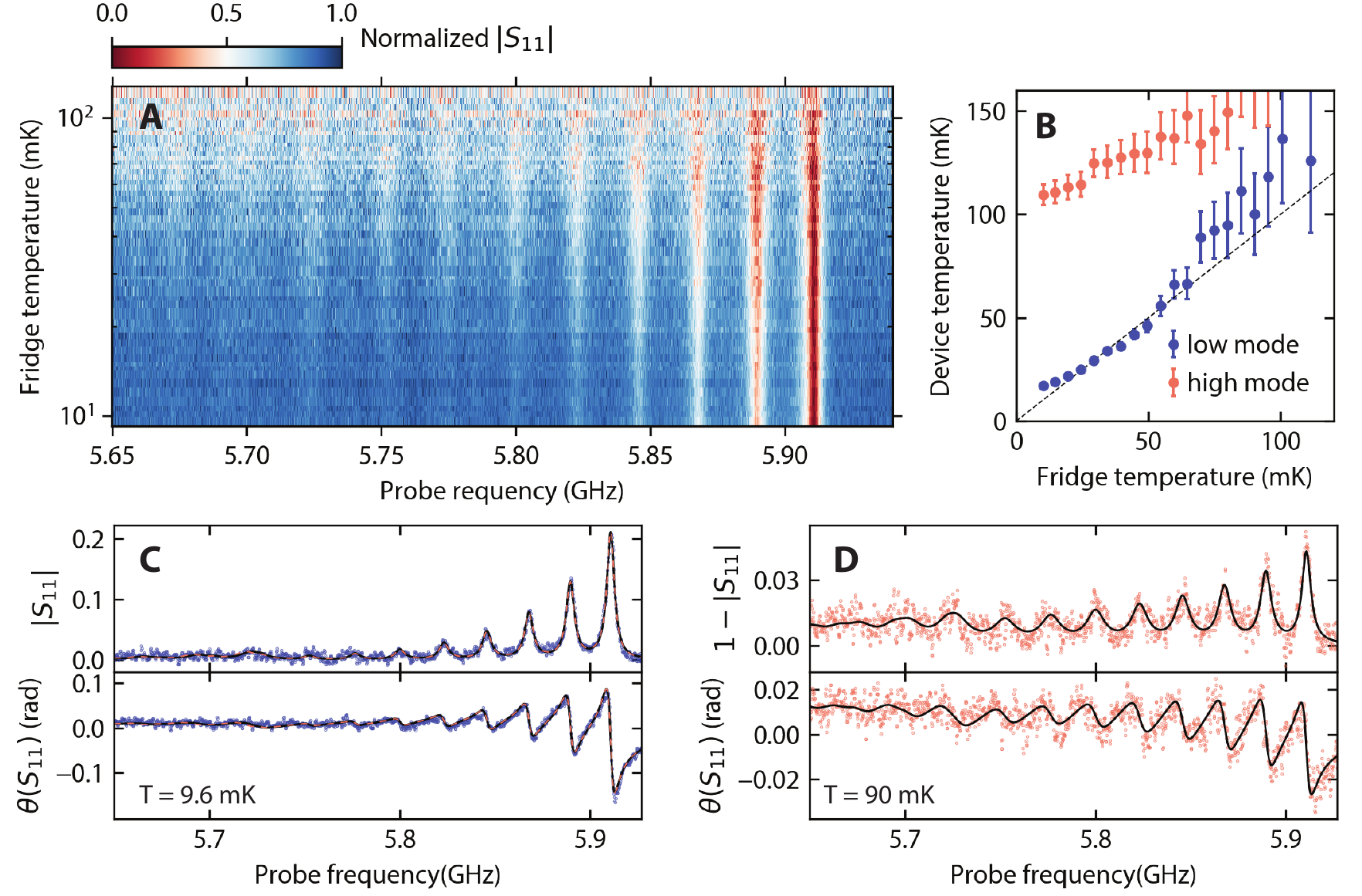}
\caption[Temperature dependence]{
	\textbf{Temperature dependence. A: }
	Normalized $|S_{11}|$ as a function of probe frequency and fridge temperature. 
	\textbf{B: }
	Temperature of both modes, fit using using Eq.~(\ref{eq:sum_of_lorentzians}), as a function of fridge temperature.
	\textbf{C: }
	Lowest-temperature data (blue) and two fits, one using Eq.~(\ref{eq:sum_of_lorentzians}) (black), and another fitting a numerical model as described in Sec.~\ref{sec:numerics_S11} (dashed red line).
	Excellent agreement between both fits validates our method of fitting the spectrum with a sum of Lorentzian functions.
	\textbf{D: }
	Higher-temperature data (red) and fit (black) using Eq.~(\ref{eq:sum_of_lorentzians}).
}
\label{fig:S_temperature}
\end{figure*}

Using $\gamma/2\pi=23\cdot 10^3s^{-1}$, we fit the spectra shown in Fig.~\ref{fig:S_temperature} to fix $\kappa$, $\kappa_\text{ext}$ and $n_\text{th}^{(H)}$.
Here, the fridge temperature is varied, and from a fit of Eq.~(\ref{eq:sum_of_lorentzians}) we extract $\kappa,\kappa_\text{ext}$, $n_\text{th}$ and $n_\text{th}^{(H)}$ at each temperature.
We took care to let the system thermalize for $\sim 10$ minutes at each temperature before starting measurements.
The linear scaling of low-mode temperature with fridge temperature, shown in Fig.~\ref{fig:S_temperature}B, confirms that we can extract a realistic mode temperature from the Bose-Einstein distribution.
A large difference in temperature is measured between low and high mode, which could be explained by the difference in external coupling to the feedline of the two modes.
We fix the values of $\kappa$, $\kappa_\text{ext}$ and $n_\text{th}^{(H)}$ to the lowest fridge temperature fit (Fig.~\ref{fig:S_temperature}C).
We leave $n_\text{th}$ as free parameters in the other experiments as it was found to vary by 10 to 20 percent on a time-scale of hours.
In the main text we quote the value of $n_\text{th}$ of the lowest point in the temperature sweep ($n_\text{th}=1.62$), but in the Fock state stabilization measurement, we measured $n_\text{th}=1.40$, in the cooling experiment $n_\text{th}=1.81$ and in the time-domain $n_\text{th}=1.37$.
These fluctuations are much smaller than the uncertainty in fitting $n_\text{th}$.
In both the cooling and Fock state stabilization experiments, $n_\text{th}$ was extracted from an initial measurement of $S_{11}$ in absence of pump tones.
%

%

\subsection{Circuit parameters}
The frequency of the system transitions (and hence the circuit parameters) is determined by fitting a numerical steady-state calculation of $S_{11}$ to the lowest temperature data (Fig.~\ref{fig:S_cooling}C).
This simulation, described in \ref{sec:numerics_S11}, starts with a diagonalization of the Hamiltonian of Eq.~(\ref{eq:Hamiltonian_8th_order}).
In this fit we additionally impose that the transition frequency $\ket{g,0}\leftrightarrow\ket{g,1}$ match the value measured in two-tone spectroscopy (Fig.~\ref{fig:S_two_tone}A).
%
%

\begin{figure*}[ht!]
\includegraphics[width=1\textwidth]{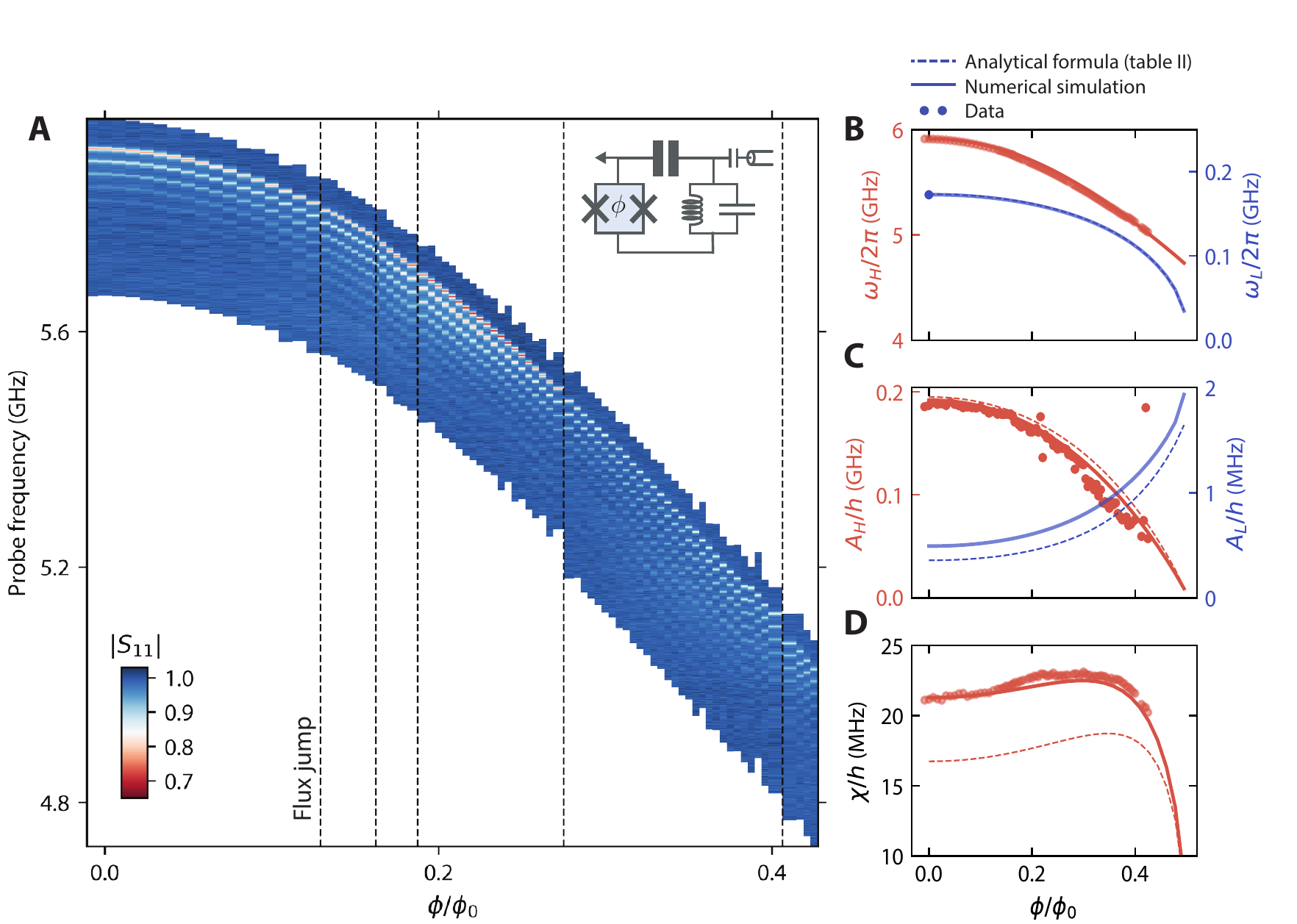}
\caption[Flux dependence of the system parameters]{
	\textbf{Flux dependence of the system parameters. A: }
	$|S_{11}|$ as a function of probe frequency and flux through the SQUID. 
	Dashed black lines correspond to flux jumps.
	\textbf{B,C,D: }
	Eigenfrequencies, anharmonicities and cross-Kerr coupling of the system as a function of flux.
	Dots are extracted through a sum-of-Lorentzians fit of the dataset in A.
	Full curves correspond to a numerical diagonalization of the Hamiltonian of Eq.~(\ref{eq:Hamiltonian_8th_order}), Taylor expanded to the 8-th order.
	Dashed lines correspond to analytical formulas obtained from black-box quantization.
	The single data point corresponding to the low-mode frequency is extracted from the sideband transition frequencies (Fig.~\ref{fig:S_two_tone}).
}
\label{fig:S_flux}
\end{figure*}

We further verify the values of $C_\text{H}$, $C_\text{L}$, $L_\text{J}$ and $L$, as well as the black-box circuit analysis of Sec.~\ref{sec:black_box}, by extracting $A_\text{H}$, $\chi$ and $\omega_\text{H}$ for a varying $L_\text{J}$.
The junction or rather SQUID inductance is modified by sweeping the flux traversing it.
This is done by current-biasing a coil situated beneath our sample.
We show in Figs.~\ref{fig:S_flux}B,C,D the result of fitting a sum of Lorentzians to the flux-dependent spectrum (Fig.~\ref{fig:S_flux}A).
For each extracted parameter, we plot the theoretical evolution with flux obtained through a numerical diagonalization of the Hamiltonian of Eq.~(\ref{eq:Hamiltonian_8th_order}) (Taylor expanded to the 8-th order), as well as the analytical expressions obtained from black-box quantization (Eqs.~(\ref{eq:wl},\ref{eq:Al},\ref{eq:wh},\ref{eq:Ah})).
The only discrepancy is between the numerical and analytical estimation of $A_\text{L}$ and $\chi$.
It arises due to a term obtained from the quartic non-linearity of the junction proportional to: $(\hat a ^\dagger \hat a +1) (\hat a ^\dagger + \hat a )(\hat b ^\dagger + \hat b ) + h.c.$. 
This term resembles a beam-splitter interaction which typically makes an oscillator more anharmonic when coupled to an oscillator more non-linear than itself.

The asymmetry of the SQUID dictating the dependence of $L_\text{J}$ on flux was a fit parameter in the construction of this figure and was found to be $20\%$.
This experiment also suffered from a number of flux jumps, where the transition frequency of the circuit suddenly jumped to a different value.
The flux was then swept until we recovered the same frequency before continuing the scan.
This data-set is thus assembled from 6 different measurements.
Therefore, an entire flux periodicity was not successfully measured, making the conversion between the current fed into a coil under the sample and flux a free parameter.

\onecolumngrid
\clearpage
{\renewcommand{\arraystretch}{1.2}
\begin{table}[]
\centering
\caption{Fitted system parameters}
\begin{tabular}{L{6cm}L{1.5cm}L{3cm}L{3cm}}
\\

Quantity                                             & Symbol              & Value     & Equation \\ \hline
\\\multicolumn{4}{c}{Hamiltonian parameters}\\\\
Dressed high-mode frequency ($\ket{g,0}\rightarrow\ket{e,0}$) & $\omega_H$          & $2\pi\times$ 5.911~GHz & $\bar\omega_H+A_H/\hbar+\chi/2\hbar$ \\
Dressed low-mode frequency ($\ket{g,0}\rightarrow\ket{g,1}$)  & $\omega_L$          & $2\pi\times$ 173~MHz & $\bar\omega_L+A_L/\hbar+\chi/2\hbar$\\
\\
Bare high-mode frequency                                  & $\bar{\omega}_H$    & $2\pi\times$ 6.113~GHz          &  $\sqrt{\frac{L+L_J}{LL_JC_L}}$        \\
Bare low-mode frequency                                   & $\bar{\omega}_L$    & $2\pi\times$ 182~MHz          &  $\frac{1}{\sqrt{(L+L_J)C_L}}$        \\ 
\\
High-mode anharmonicity                               & $A_H$               &  $h\times$ 192~MHz  &  $\frac{e^2}{2C_H}\left(\frac{L}{L+L_J}\right)$,         \\
Low-mode anharmonicity                                & $A_L$               &  $h\times$ 495 kHz  &  $\frac{e^2}{2C_L}\left(\frac{L_J}{L+L_J}\right)^3$,       \\
Cross-Kerr                                           & $\chi$              &  $h\times$ 21.29~MHz   &  $2\sqrt{A_LA_H}$        \\
\\\multicolumn{4}{c}{Dissipation rates}\\\\
High-mode dissipation rate                            & $\kappa$            &  $2\pi\times$ 3.70~MHz         &  \\
External coupling rate                               & $\kappa_\text{ext}$          &  $2\pi\times$ 1.63~MHz         &         \\
Low-mode dissipation rate                             & $\gamma$            &  $2\pi\times$ 23.50 kHz         &         \\
Low-mode external coupling rate                       & $\gamma_\text{ext}$          &  $2\pi\times$ 1.99 Hz         &        \\
\\
High-mode quality factor                              & $Q_H$               &  1599         &  $\omega_H/\kappa$        \\
High-mode external quality factor                     & $Q_H^{(\text{ext})}$         &  3617         &  $\omega_H/\kappa_\text{ext}$        \\
Low-mode quality factor                               & $Q_L$               &  7348         &  $\omega_L/\gamma$        \\
Low-mode external quality factor                      & $Q_L^{(\text{ext})}$         &  87 $\times 10^6$        &  $Z_0\sqrt{\frac{C_L}{L+L_J}}\left(\frac{C_c}{C_L}\right)^2$        \\
\\\multicolumn{4}{c}{Thermal parameters}\\\\
High-mode temperature                                 & $T_H$               &  112~mK         &   \\
Low-mode temperature                                  & $T_L$               &  17~mK         &    \\
\\
High-mode occupation number                           & $n_\text{th}^{(H)}$ &  0.09         &  $\frac{1}{e^{\frac{\hbar\omega_H}{k_BT_H}}-1}$        \\
Low-mode occupation number                            & $n_\text{th}$       &  1.62        &  $\frac{1}{e^{\frac{\hbar\omega_L}{k_BT_L}}-1}$          
\\\multicolumn{4}{c}{Circuit parameters}\\\\
Josephson energy                                     & $E_J$               & $h\times$ 4.01~GHz &  $\frac{\hbar^2\bar{\omega}_H^2\bar{\omega}_L^2}{8\left(\bar{\omega}_H\sqrt{A_L}+\bar{\omega}_L\sqrt{A_H}\right)^2}$        \\
Josephson inductance                                 & $L_J$               & 41 nH          &  $\frac{\hbar^2}{4e^2E_J}$        \\
Low-mode capacitance                                  & $C_L$               & 11.1 pF          &  $\frac{e^2 \sqrt{A_L}  \bar{\omega}_H^3}{2 \left(\bar{\omega}_L\sqrt{A_H}+\bar{\omega}_H\sqrt{A_L}\right)^3}$       \\
High-mode capacitance                                 & $C_H$               & 40.7~fF          &  $\frac{e^2 \bar{\omega}_L}{2 \left(\sqrt{A_HA_L} \bar{\omega}_H+A_H\bar{\omega}_L\right)}$        \\
High-mode inductance                                  & $L$                 & 28.2 nH          &  $\frac{2 \sqrt{A_H} \left(\bar{\omega}_L\sqrt{A_H}+\bar{\omega}_H\sqrt{A_L}\right)^2}{e^2\sqrt{A_L}  \bar{\omega}_H^3 \bar{\omega}_L}$        \\
Coupling capacitor                                   & $C_c$               & 0.95~fF          &  $C_H\sqrt{\frac{\kappa_\text{ext}LL_J}{Z_0(L+L_J))}}   $    \\
Feedline impedance                                   & $Z_0$               & 50$\Omega$          &        \\

\end{tabular}
\label{tab:params}
\end{table}
}
\pagebreak
\FloatBarrier
\twocolumngrid

\onecolumngrid
\section{Supplementary experimental data}

\subsection{Flux dependence of thermal and dissipation parameters}

\begin{figure*}[h!]
\includegraphics[width=0.8\textwidth]{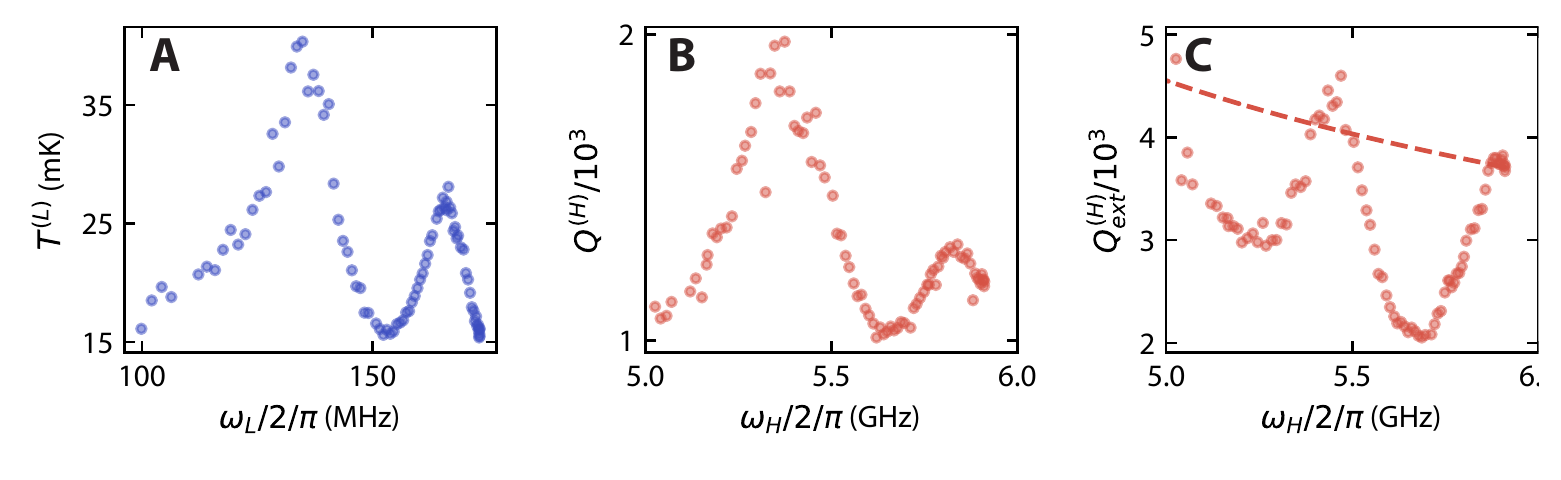}
\caption[Flux dependence of thermal and dissipation parameters]{
	\textbf{Flux dependence of thermal and dissipation parameters. A: }
	Low mode temperature, as a function of the low-mode resonance frequency.
	\textbf{B,C: }Total and external quality factor of the high mode as a function of its frequency.
	The dashed line in panel C corresponds to the expected behavior from our circuit analysis.
	These parameters are extracted from a fit of Eq.~(\ref{eq:sum_of_lorentzians}) to the flux-dependent spectrum shown in Fig.~\ref{fig:S_flux}.
}
\label{fig:S_flux_n_th_kappa}
\end{figure*}

\begin{figure*}[ht!]

\subsection{Full time-dependent spectrum}
\includegraphics[width=0.8\textwidth]{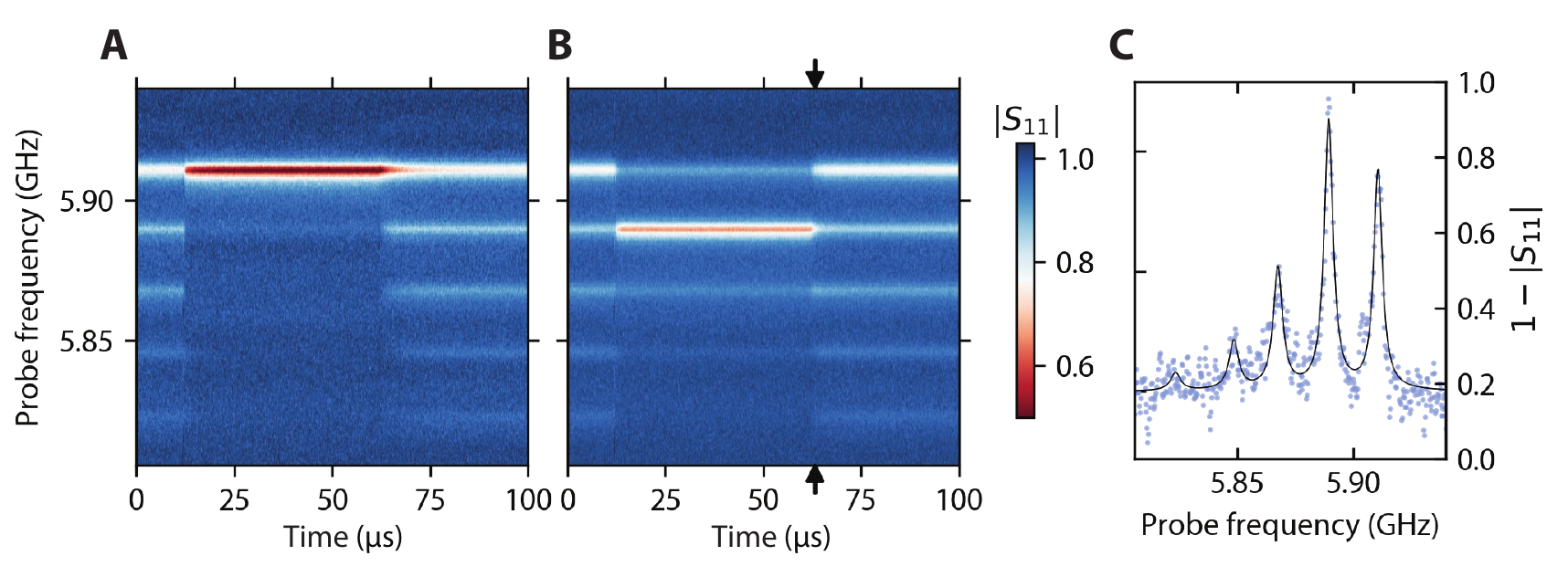}
\caption[Full time-dependent spectrum]{\textbf{Full time-dependent spectrum. }
	Time and probe frequency dependence of $|S_{11}|$ for both ground-state cooling (\textbf{A}) and the one-photon-state stabilization (\textbf{B}).
	By fitting these datasets using Eq.~(\ref{eq:sum_of_lorentzians}), in both frequency and time, we construct the plots shown in Fig.~4.
	\textbf{C: }line cut of the data set B (indicated by arrows in B) is shown as blue dots, the black line corresponds to a fit.
	The relatively low signal-to-noise ratio is responsible for the large noise in probability of Fig.~4.
}
\label{fig:S_time_domain}
\end{figure*}

The flux sweep shown in Fig.~\ref{fig:S_flux} also gives access to the temperature and dissipation rates of the modes as a function of flux which are shown in Fig.~\ref{fig:S_flux_n_th_kappa}.
These are extracted from a fit of the sum of Lorentzians (Eq.~(\ref{eq:sum_of_lorentzians})), \textit{i.e.} from the line-widths and amplitudes of the measured peaks.
This relies on our estimation of $\gamma$, which is assumed to be a constant, and $n_\text{th}^{(H)}$, which is difficult to extract due to the low signal-to-noise ratio as well as the $\ket{e,0}\leftrightarrow\ket{f,0}$ peak crossing the $\ket{g,n}\leftrightarrow\ket{e,n}$ peaks.
To investigate the accuracy of these fits, we plot in Fig.~\ref{fig:S_flux_n_th_kappa}C the external quality factor of the high mode as a function of its frequency (Fig.~\ref{fig:S_flux_n_th_kappa}C) .
We find a clear mismatch with the behavior expected from our circuit analysis.
This indicates that we cannot confidently state that the temperature of the low mode and dissipation of the high mode fluctuate with flux as shown here, or even provide meaningful error bars.
Further analysis could take the form of time-domain measurements at each flux points to determine $\gamma$, or higher signal-to-noise measurements of the $\ket{e,0}\leftrightarrow\ket{f,0}$ to fix $n_\text{th}^{(H)}$.

\FloatBarrier
\subsection{Four-wave mixing spectrum}

\begin{figure*}[ht!]
\includegraphics[width = 0.7\columnwidth]{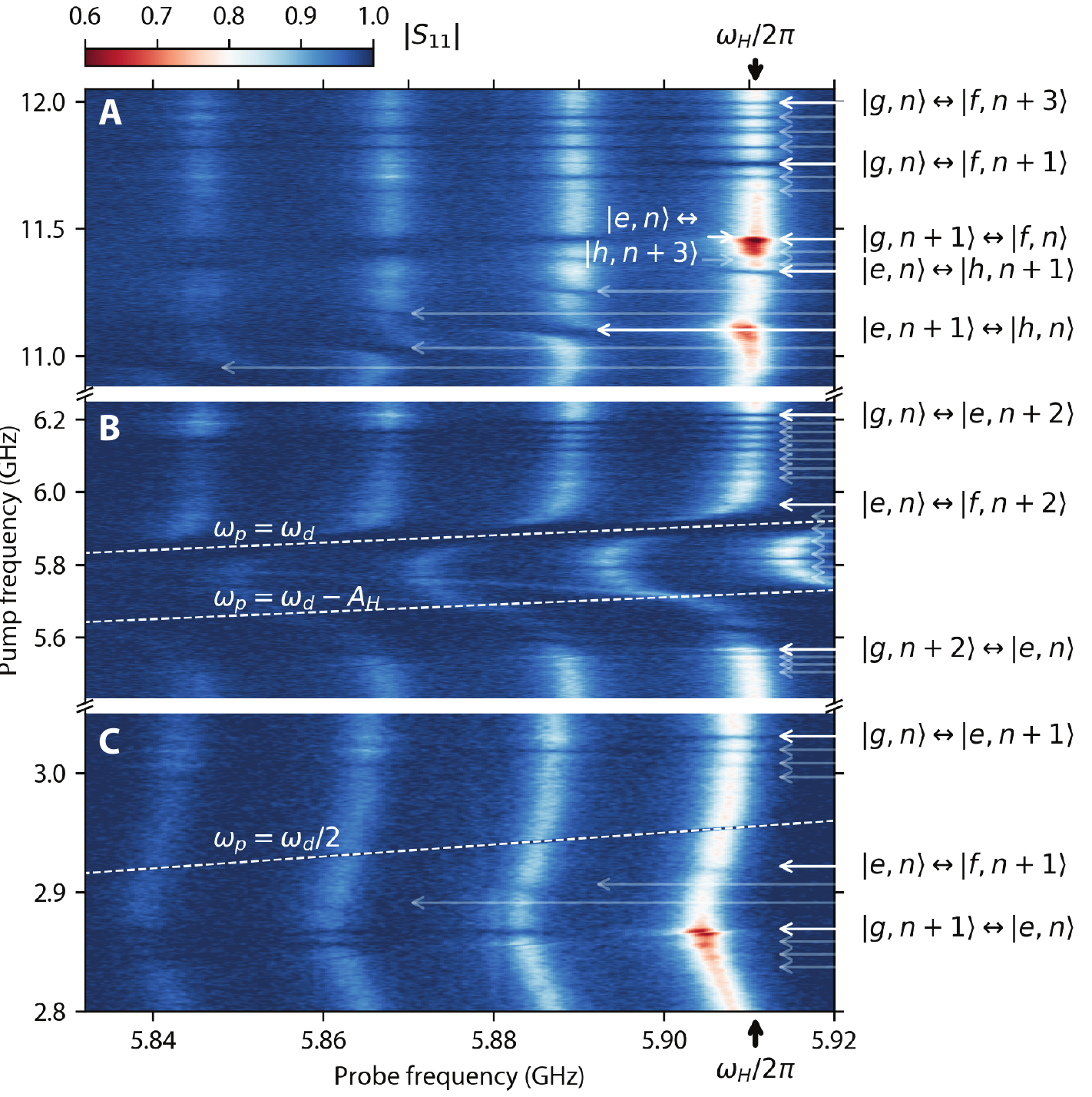}
\caption[Measurement of four-wave mixing processes]{
\textbf{Measurement of four-wave mixing processes.}
	$|S_{11}|$ as a function of probe frequency and of the frequency of a stronger pump tone.
	Features in the data are indicated by arrows with position generated from the eigenvalues of the Hamiltonian of Eq.~(\ref{eq:Hamiltonian_8th_order}), Taylor expanded to the 8-th order.
	%
	%
	The AC-Stark shift is not considered in the computation of transition energies, and a constant high-mode frequency is taken as indicated by a black arrow in the x-axis, leading to slight mismatches between the transitions and placed arrows.
	In panel \textbf{B} dashed lines indicate the large avoided crossings observed when the pump tone is directly resonant with the transition frequencies of the high mode.
	In panel \textbf{C}, the dashed line indicates $\omega_p=\omega_d/2$, the features arising there being due to the first harmonic overtone of the pump (issuing from our microwave generator) driving the high mode.
}
\label{fig:S_all_transitions}
\end{figure*}

By measuring the spectrum whilst sweeping the frequency of a pump tone, we show in Fig.~\ref{fig:S_all_transitions} the multitude of four-wave mixing processes possible in this system.
Panel A is particularly relevant to the cooling experiment, and is shown in Fig.~3A, as one can see the relevant transitions lying next to the cooling transition $\ket{g,0}\leftrightarrow\ket{f,1}$.
We tested different combinations of raising and cooling four-wave mixing processes (panels B and C) for cooling and Fock-state stabilization, but these alternatives consistently produced lower state occupations than the results shown in the main text.

Two transitions in panel A are unexpected from a simple four-wave mixing approach to the system: $\ket{g,n}\leftrightarrow\ket{f,n+3}$ and $\ket{e,0}\leftrightarrow\ket{h,n+3}$.
These are six-wave mixing processes and one could expect them to have very weak effects.
However in this system the cross-Kerr $\chi$ is a considerable fraction of $\omega_L$.
The usually neglected term of the quartic non-linearity of the junction proportional to $\chi(2\hat a ^\dagger \hat a +1)(\hat b\hat b+\hat b^\dagger\hat b^\dagger)$ then leads to the dressed low Fock state $\ket{n}$ having a significant overlap with the bare states $\ket{n\pm 2k}$ where k is a positive integer.
The transition $\ket{g,0}\leftrightarrow\ket{f,3}$ is thus visible since $\ket{f,3}$ has a large overlap with $\ket{f,1}$ and $\ket{g,0}\leftrightarrow\ket{f,1}$ is an easily drivable four-wave mixing transition.

\subsection{Low-frequency spectrum}

\begin{figure*}[ht!]
\includegraphics[width = 0.8\columnwidth]{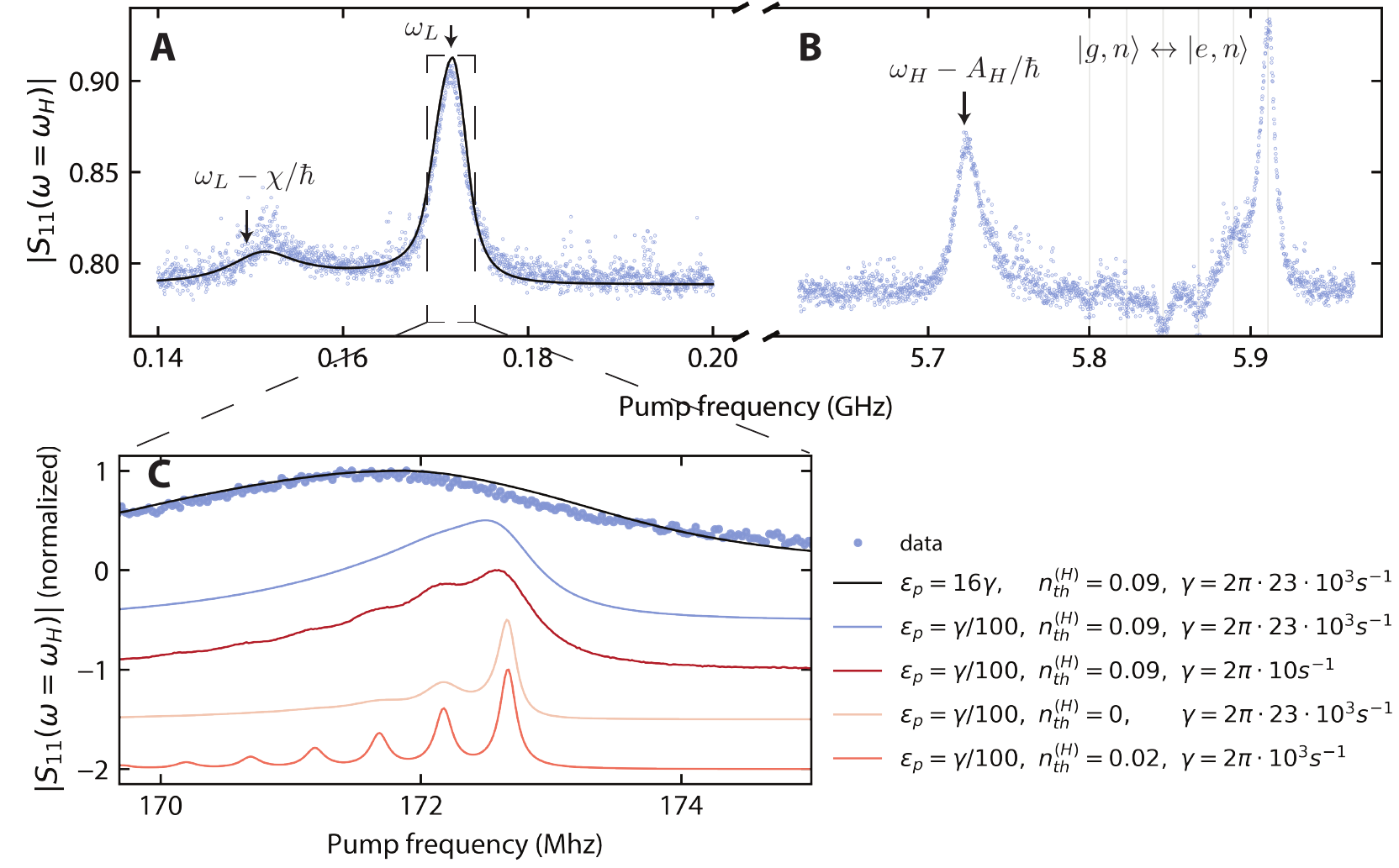}
\caption[Two-tone measurement of the anharmonicity and lower mode spectrum]{\textbf{Two-tone measurement of the anharmonicity and lower mode spectrum. }
\textbf{A: }we sweep a pump tone around the low-mode frequency (x-axis) whilst monitoring the depth of the $n=0$ peak $|S_{11}(\omega=\omega_\text{H})|$ (y-axis).
We observe two peaks, separated by $\chi$ corresponding to the transitions $\ket{g,n}\leftrightarrow\ket{g,n+1}$ at $\omega_\text{p}=\omega_\text{L}$ and $\ket{e,n}\leftrightarrow\ket{e,n+1}$ at $\omega_\text{p}=\omega_\text{L}-\chi/\hbar$.
A steady-state numerical computation is shown as a black line and data as blue points.
\textbf{B: }by performing the same measurement around the high-mode frequency, we measure a peak corresponding to $\ket{e,0}\leftrightarrow\ket{f,0}$ at $\omega_\text{p}=\omega_\text{H}-A_\text{H}/\hbar$.
Compared to Fig.~\ref{fig:S_all_transitions}, these two datasets constitute a more direct measurement of $\omega_\text{L}$ and $A_\text{H}$.
\textbf{C: }the line-width of the low mode was found to be significantly broader than $\gamma$, with no accessible photon-number resolution.
By varying simulation parameters as detailed in the legend, we explore the origin of this broad line-width.
}
\label{fig:S_two_tone}
\end{figure*}

We monitor the height of the $\ket{g,0}\leftrightarrow\ket{e,0}$, whilst sweeping the frequency of a secondary pump tone.
As shown in Fig.~\ref{fig:S_two_tone}A,B, this allows us to easily measure the anharmonicity of the high mode and the frequency of the low mode.
The line-width of the low-mode peak is considerably larger than the previously determined low-mode dissipation rate $\gamma/2\pi=23\cdot 10^3s^{-1}$.
If the line-width was equal to $\gamma$, we would expect to see photon number splitting, distinct peaks separated by the low-mode anharmonicity $A_\text{L}$, corresponding to the transitions $\ket{g,n}\leftrightarrow\ket{g,n+1}$.
To understand why this is not the case, we fit a steady-state numerical computation of a pumped and probed Hamiltonian 
\begin{equation}
\begin{split}
	\hat H&=-A_\text{H}(\hat a^\dagger)^2\hat a^2 +\hbar(\omega_\text{L}-\omega_\text{p})\hat b^\dagger\hat b -A_\text{L}(\hat b^\dagger)^2\hat b^2\\ 
	&+ i\hbar\epsilon_\text{d}(\hat a^\dagger-\hat a)+ i\hbar\epsilon_\text{p}(\hat b^\dagger-\hat b)\ ,
\end{split}
\end{equation}
with the collapse operators of Eq.~(\ref{eq:probed_lindbald}).
The only free parameter is the pumping strength $\epsilon_\text{p}\sim16\times\gamma$, the probe strength was taken to be negligibly small with respect to all other rates in the model.
By varying simulation parameters, we can then explore the origin of this broad line-width. 
These results are summarized in Fig.~\ref{fig:S_two_tone}C.
Reducing the pumping strength $\epsilon_\text{p}$ will suppress what is usually referred to as `power broadening', at the expense of the signal-to-noise ratio, but does not reveal photon-number splitting.
By reducing $\gamma$ to a negligibly small rate, photon number splitting can only be glimpsed behind a line-width broadening induced by the process $\ket{g,n}\rightarrow\ket{e,n}$ which occurs at a rate $\kappa n_\text{th}^{(H)}$.
This becomes clear if we instead keep $\gamma/2\pi=23\cdot 10^3s^{-1}$ and take the limit $n_\text{th}^{(H)}=0$, making the first two peaks apparent. 
As derived in Eq.~(\ref{eq:sum_of_lorentzians}), the line-width of a thermally populated anharmonic oscillator broadens significantly with its thermal occupation, which is responsible in this case for the disappearance (broadening) of peaks $n\ge2$.
By reducing both $\gamma$ and $n_\text{th}^{(H)}$, photon-number resolution would become visible.

\end{document}